\documentclass[11pt,a4paper]{article}
\usepackage{jheppub,amsmath, amsthm, amssymb,slashed,url}
\usepackage{graphicx}
\usepackage{epstopdf}
\usepackage[all]{xy}
\usepackage{multirow}

\font\teneurm=eurm10 \font\seveneurm=eurm7 \font\fiveeurm=eurm5
\newfam\eurmfam
\textfont\eurmfam=\teneurm \scriptfont\eurmfam=\seveneurm
\scriptscriptfont\eurmfam=\fiveeurm

\font\teneusm=eusm10 \font\seveneusm=eusm7 \font\fiveeusm=eusm5
\newfam\eusmfam
\textfont\eusmfam=\teneusm \scriptfont\eusmfam=\seveneusm
\scriptscriptfont\eusmfam=\fiveeusm

\font\tencmmib=cmmib10 \skewchar\tencmmib='177
\font\sevencmmib=cmmib7 \skewchar\sevencmmib='177
\font\fivecmmib=cmmib5 \skewchar\fivecmmib='177
\newfam\cmmibfam
\textfont\cmmibfam=\tencmmib \scriptfont\cmmibfam=\sevencmmib
\scriptscriptfont\cmmibfam=\fivecmmib
\def\pa{\partial}

\newcommand{\mc}[1]{\mathcal{#1}}

\newcommand{\si}{\sigma}
\newcommand{\lam}{\lambda}
\newcommand{\olam}{\overline{\lambda}}
\newcommand{\g}{\gamma}

\newcommand{\ophi}{\overline{\phi}}
\newcommand{\opsi}{\overline{\psi}}

\newcommand{\Sp}{\mathbb{S}}
\newcommand{\Rp}{\mathbb{RP}}
\newcommand{\compe}{\vartheta}
\newcommand{\compt}{\varphi}
\newcommand{\compd}{y}

\newcommand{\ch}{\text{\boldmath $q$}}

\newcommand{\column}[2]{
\begin{pmatrix}
#1
\\
#2
\end{pmatrix}
}

\renewcommand{\include}[1]{}
\renewcommand\documentclass[2][]{}

\title{Abelian 3d mirror symmetry on $\mathbb{RP}^2 \times \mathbb{S}^1$ with $N_f=1$}

\preprint{\begin{flushright} RIKEN-STAMP-8 \\ OU-HET 862 \end{flushright}}

\author[a]{Akinori Tanaka,}
\author[b]{Hironori Mori,}
\author[c]{and Takeshi Morita}

\affiliation[a]{iTHES Research Group, RIKEN, Wako, Saitama \\351-0198, Japan}
\affiliation[b]{Department of Physics, Graduate School of Science, Osaka University, Toyonaka, Osaka \\560-0043, Japan}
\affiliation[c]{Graduate School of Information Science and Technology, Osaka University, Toyonaka, Osaka \\560-0043, Japan}

\abstract{
We consider a new 3d superconformal index defined as the path integral over $\Rp^2 \times \Sp^1$, and get the generic formula for this index with arbitrary number of U$(1)$ gauge symmetries via the localization technique.
We find two consistent parity conditions for the vector multiplet, and  name them $\mc{P}$ and $\mc{CP}$.
We find an interesting phenomenon that
two matter multiplets coupled to the $\mc{CP}$-type vector multiplet  merge together.
By using this effect, we investigate the simplest version of 3d mirror symmetry on $\Rp^2 \times \Sp^1$ and observe four types of coincidence between the SQED and the XYZ model.
We find that merging two matters plays a important role for the agreement.
}
\begin{document} \maketitle


\section{Introduction}
The SUSY localization technique on a \textit{curved manifold} $\mc{M}$ provides us an exact way to perform the path integral.
For example, exact results based on the localization techniques on the $3$-dimensional manifold $\mc{M}_3$ have been reported in early works \cite{Kapustin:2009kz, Kim:2009wb, Gang:2009wy, Imamura:2011su}.
Once we know such a manifold, we can study possible 
orbifold \cite{Imamura:2012rq} and the boundary condition \cite{Sugishita:2013jca} preserving SUSY.
Another \text{exotic} story can be found in the study of rigid SUSYs on an \textit{unorientable} manifold \cite{Tanaka:2014oda}. 
We can observe different aspects of SUSY gauge theory by considering its partition function on various manifolds.

As one of interesting applications of these exact results, we focus on the simplest version of 3d mirror symmetry \cite{Intriligator:1996ex, Aharony:1997bx} which is equivalence between the SQED and the XYZ model.
In fact, 3d mirror symmetry can be embedded to the context of string theory, and mirror theories are related by SL$(2,\mathbb{Z})$ transformation in the type IIB brane construction \cite{deBoer:1996mp}.
By using exact results on 
%
$\mc{M}_3 = \Sp^3$, we can find the expected agreement of partition functions via the Fourier transform of hyperbolic function sech, or 
the canonical transformation of a certain phase space \cite{Drukker:2015awa}.
If we consider partition functions on the squashed three sphere \cite{Hama:2011ea}, it is generalized to the Fourier transform of the double sine function \cite{Hama:2010av} and becomes the summation identity by considering $\mc{M}_3 = \Sp^3/\mathbb{Z}_p$ \cite{Imamura:2012rq}.
If we take $\mc{M}_3 = \Sp^2 \times \Sp^1$, the expected identity is proved by using non-trivial identities called the $q$-binomial formula and the Ramanujan's summation formula \cite{Krattenthaler:2011da, Kapustin:2011jm}.
These results show us that there is a complementary relationship between a family of mathematical identities depending on $\mc{M}_3$ and the validity of the original 3d mirror symmetry.

\newpage

In \cite{Tanaka:2014oda}, we studied the case of $\mc{M}_3 = \Rp^2 \times \Sp^1$. 
We realized the theory on such an \textit{unorientable space} via dividing $\Sp^2 \times \Sp^1$ by $\mathbb{Z}_2$ parity $\mc{P}$ which acts on the base space $\Sp^2$, and computed the superconformal index.
One important limitation of $\mc{P}$ is compatibility with SUSY because we utilize the supersymmetric localization technique.
The simplest case of 3d mirror symmetry is realized as equivalence between the following quiver diagrams:
\begin{align}
\xymatrix {
*+[F-]{Q}
\ar@{->}[r]
&
*++[o][F-]{\text{\scriptsize$\mc{P}$}} 
\ar@{->}[r]
&
*+[F-]{\tilde{Q}}
}
\quad
\Longleftrightarrow
\quad
\xymatrix {
*+[F-]{\text{\scriptsize$
\begin{matrix}
X \\
Y
\end{matrix}
$}
}
&
*+[F-]{Z}
}
\end{align}
The symbol $\mc{P}$ in the node means that the gauge field $A_\mu$ in the vector multiplet transforms like $\pa_\mu$ after the translation along the non-contractible cycle on $\Rp^2$, and the combined $X,Y$ means that $X$ transforms into $Y$, and vice versa.
We can verify equivalence between two indices from the $q$-binomial theorem and a parity formula for $q$-Pochhammer symbol:
\begin{align} 
&
q^{\frac{1}{8}}
\frac{( q^{2} ; q^{2} )_{\infty}}{( q ; q^{2} )_{\infty}}
\oint
\frac{d z}{2 \pi i z} 
\left\{
a^{- \frac{1}{4}}
\frac
{( z^{- 1} a^{\frac{1}{2}} q^{\frac{1}{2}}, z a^{\frac{1}{2}} q^{\frac{1}{2}}; q^{2} )_{\infty}}
{( z a^{- \frac{1}{2}} q^{\frac{1}{2}}, z^{- 1} a^{- \frac{1}{2}} q^{\frac{1}{2}} ; q^{2} )_{\infty}}
+
a^{\frac{1}{4}}
\frac
{( z^{- 1} a^{\frac{1}{2}} q^{\frac{3}{2}}, z a^{\frac{1}{2}} q^{\frac{3}{2}} ; q^{2} )_{\infty}}
{( z a^{- \frac{1}{2}} q^{\frac{3}{2}}, z^{- 1} a^{- \frac{1}{2}} q^{\frac{3}{2}} ; q^{2} )_{\infty}} \right\} 
\notag \\
&\Leftrightarrow
q^{\frac{1}{8}} \tilde{a}^{- \frac{1}{4}}
\frac{( \tilde{a}^{- \frac{1}{2}} q^{} ; q )_{\infty}}{( \tilde{a}^{\frac{1}{2}} 
; q )_{\infty}}
\frac{( \tilde{a} ; q^{2} )_{\infty}}{( \tilde{a}^{- 1} q ; q^{2} )_{\infty}}.
\end{align}

\vspace{.5cm}
In this paper, we find a pair of new consistent parity conditions 
by turning on \textit{charge conjugation} $\mc{C}$ simultaneously with $\mc{P}$, and derive new superconformal indices for a general class of theories based on localization techniques.
We find non-trivial effects on the structure of localization locus and the one-loop determinants, and utilize the new results to the check of 3d mirror symmetry with one flavor.
We find following expected equivalence:
\begin{align}
\xymatrix {
*+[F-]{\text{\scriptsize$
\begin{matrix}
Q \\
\tilde{Q}
\end{matrix}
$}
}
\ar@{->}[r]<1mm>
\ar@{<-}[r]<-1mm>
&
*++[o][F-]{\text{\scriptsize$\mc{CP}$}} 
}
\quad
\Longleftrightarrow
\quad
\xymatrix {
*+[F-]{X}
&
*+[F-]{Y} 
&
*+[F-]{Z}
}
\end{align}
The symbol $\mc{CP}$ in the node means that the gauge field $A_\mu$ transforms oppositely compared with $\pa_\mu$ along the non-contractible cycle. 
We can verify corresponding equivalence between two indices by using the Ramanujan's summation formula and the product-to-sum identity of $\vartheta$ functions:
\begin{align} 
&
\frac{1}{2}
 q^{- \frac{1}{8}}
\frac{( q; q^{2} )_{\infty}}{( q^{2}; q^{2} )_{\infty}}
\sum_{m \in \mathbb{Z}}
\left( q^{\frac{1}{2}} a ^{- \frac{1}{2}} w \right)^{m}
\Bigg\{
\frac{( a^{- \frac{1}{2}} q^{m + 1}; q )_{\infty}}{( a^{\frac{1}{2}} q^{m}; q )_{\infty}}
+
\frac{( - a^{- \frac{1}{2}} q^{m + 1}; q )_{\infty}}{( - a^{\frac{1}{2}} q^{m}; q )_{\infty}}
\Bigg\}
\notag \\ & 
\Leftrightarrow
q^{- \frac{1}{8}}
\frac
{( \tilde{a}^{\frac{1}{2}} \tilde{w}^{- 1} q^{\frac{1}{2}}, \tilde{a}^{\frac{1}{2}} \tilde{w} q^{\frac{1}{2}}, \tilde{a}^{- 1} q; q^{2} )_{\infty}}
{( \tilde{a}^{- \frac{1}{2}} \tilde{w} q^{\frac{1}{2}}, \tilde{a}^{- \frac{1}{2}} \tilde{w}^{- 1} q^{\frac{1}{2}}, \tilde{a}; q^{2} )_{\infty}}.
\end{align}
We also consider the generalized index and observe four types of perfect agreement between the SQED and the XYZ model.

The organization of this paper is as follows.
In Section 2, we explain details on the new parity conditions and summarize the localization results.
In Section 3, we turn to showing four types of coincidence of two generalized indices conjectured by the 3d mirror symmetry.
Section 4 is devoted to the conclusion and discussions.
We add three appendices. In Appendix \ref{LOCALIZATION}, we explain more details on localization calculus.
In Appendix \ref{24mirrorsym}, we summarize useful mathematical formulas and give analytic proof for four types of coincidence.

\bibliographystyle{common_supp_files/sonota/utphys}
\bibliography{common_supp_files/Ref}


\section{Exact computations of 3d superconformal index on $\Rp^2 \times \Sp^1$}
\paragraph{Preliminary}
Our basic ingredients for constructing SUSY gauge theories are a 3d $\mc{N}=2$ vector multiplet $V$ with a U$(1)$ gauge symmetry and a matter multiplet $\Phi$ on $\Sp^2 \times \Sp^1$.
Under the $\mc{N} = 2$ supersymmetry, the following Lagrangians define SUSY-invariant actions: 
\begin{align}
	& \fbox{\text{{Lagrangians}}} \notag
	\\
&
\mc{L}_{\text{YM}} (V)
=
 \frac{1}{2}    F_{\mu \nu}  F^{\mu \nu}
 + D^2
 +\pa_\mu \si \cdot \pa^\mu \si
 + 
 \epsilon^{3 \rho \si} \si F_{\rho \si} 
 + 
 \si^2 
+ i \olam \g^\mu \nabla_\mu \lam
- \frac{i}{2} \olam \g_3 \lam
,
\label{YM}
%
\\
&
\mc{L}_{\text{mat}} (\Phi ; \ch) 
=
-i   \opsi \g^\mu (\nabla_\mu - i \ch A_\mu) \psi
+    ( \pa_\mu  \ophi  + i \ch   \ophi  A_\mu)
( \pa^\mu \phi   - i \ch A^\mu \phi  ) 
+\cdots
.
\label{mat}
\end{align}
These Lagrangians are not only SUSY-invariant but also SUSY-\textit{exact}; we can rewrite them as consequences of the SUSY variations. (See \cite{Tanaka:2014oda} for more details.)
Thanks to their exactness, we can use the localization technique.
In addition, we should take appropriate parity conditions to component fields in order to make the theory defined on $\Rp^2 \times \Sp^1$.
We denote such parity conditions by
\begin{align}
\phi &\to  \tilde{\phi}
\
\overset{\text{def}}{\Longleftrightarrow}
\
\phi (\pi - \compe , \pi + \compt, \compd) =  \tilde{\phi} (\compe, \compt , \compd).
\end{align}
In \cite{Tanaka:2014oda}, we found a pair of parity conditions compatible with SUSY as follows.
Here, we restrict ourselves to consider only the single matter multiplet.
\\
\begin{minipage}{0.5\hsize}
  \begin{align}
  \begin{aligned}
  &\text{\fbox{{Vector multiplet} : $V^{(\mc{P})}$}}
  \\
  &A_\compe^{(\mc{P})} \to - A_\compe^{(\mc{P})},
  \quad
  A_{\compt, \compd}^{(\mc{P})} \to + A_{\compt, \compd}^{(\mc{P})},
   \\
   &\si^{(\mc{P})} \to -\si^{(\mc{P})} , 
\\
  &\lam^{(\mc{P})} \to +i \g_1 \lam^{(\mc{P})} ,
  \quad
  \olam^{(\mc{P})} \to - i \g_1 \olam^{(\mc{P})} , \\
   & 
     D^{(\mc{P})}\to+D^{(\mc{P})},
  \end{aligned}
  \label{vecp1}
  \end{align}
\end{minipage}
\begin{minipage}{0.5\hsize}
\begin{align}
	\begin{aligned}
	&\fbox{\text{{Matter multiplet} : $\Phi^{}_{s}$}}
	\\
	&
	\qquad
	\phi^{} \to  {\phi}^{},
	\\ 
	&\qquad
	{\psi}^{} \to - i \g_1  {\psi}^{},
	\\ 
	&\qquad
	{F}^{} \to   {F}^{},
	\\
	&\qquad
	\text{with a U$(1)$ charge }\ch.
	\end{aligned}
\label{ematp1}
\end{align}
\end{minipage}

\vspace{0.5cm}\noindent
Basically, these conditions are controlled by behavior of the Killing spinors on $\Sp^2 \times \Sp^1$.
In order to perform the localization calculus, we need also to make the action itself invariant under these parity transformations, and happily each Lagrangian in \eqref{YM} and \eqref{mat} is invariant independently under the transformations in \eqref{vecp1} and \eqref{ematp1} schematically described as follows:
\begin{align}
&
\xymatrix{
\mc{L}\ar @(ul,dl)}
\hspace{-1mm}
_{\text{YM}}(V^{(\mc{P})})+
\mc{L}_{\text{mat}}
(\Phi^{}_{s} ; \ch
\xymatrix{)  \ar @(ur,dr)}
\qquad
.
\label{YM_mat}
\end{align}
\paragraph{Chern-Simons term and BF term}
One may want to consider a Chern-Simons term, however it breaks parity invariance because it effectively works as the flip of the sign for the Chern-Simons level $k \to -k$.
If one turn on \textit{two} vector multiplets and do an appropriate exchange of the fields under the antipodal identification, it will be possible to preserve parity invariance, but we postpone such generic setup for a future problem.
In the same reason, we cannot turn on a supersymmetric BF term with only the conditions \eqref{vecp1}.
However, as we will see in the next page, there is another consistent parity condition \eqref{vecp2}.
Then, it becomes possible to define the BF term consistently even on $\Rp^2 \times \Sp^1$.

\newpage
\subsection{New supersymmetric parity conditions and BF term}
\label{NewCP}
\paragraph{New parity condition}
In the previous parity conditions \eqref{vecp1} and \eqref{ematp1}, we focus on a single matter multiplet.
More appropriately speaking, we consider a single $U(1)$ charge $\ch$.
In this case, we should fix the condition for the vector multiplet as \eqref{vecp1} in order to make the matter Lagrangian \eqref{mat} invariant under the transformation.
Therefore, it looks impossible to take another parity condition for the vector multiplet.
However, a loophole can be found by turning on two matter multiplets with opposite charges $+\ch, -\ch$ as follows:
\\
\begin{minipage}{0.5\hsize}
  \begin{align}
  \begin{aligned}
&\text{\fbox{{Vector multiplet} : $V^{(\mc{CP})}$}}
  \\
  &A_\compe^{(\mc{CP})} \to + A_\compe^{(\mc{CP})},
  \quad
  A_{\compt, \compd}^{(\mc{CP})} \to - A_{\compt, \compd}^{(\mc{CP})},
   \\
   &\si^{(\mc{CP})} \to +\si^{(\mc{CP})} , 
\\
  &\lam^{(\mc{CP})} \to -i \g_1 \lam^{(\mc{CP})} ,
  \quad
  \olam^{(\mc{CP})} \to + i \g_1 \olam^{(\mc{CP})} , \\
   & 
     D^{(\mc{CP})}\to-D^{(\mc{CP})}.
  \end{aligned}
  \label{vecp2}
  \end{align}
\end{minipage}
\begin{minipage}{0.4\hsize}
\begin{align}
	\begin{aligned}
	&\fbox{\text{{Matter multiplets} : $\Phi^{}_d$} }
	\\
	&\column{\phi_1}{\phi_2} \to   \column{\phi_2}{\phi_1} ,
	\column{\psi_1}{\psi_2} \to  -\column{ i \g_1\psi_2}{i \g_1\psi_1} ,
	\\ 
	&\column{F_1}{F_2} \to \column{ F_2}{F_1},
	\quad
	\text{with charges }
	\column{+\ch}{-\ch}.
	\end{aligned}
\label{ematp2}
\end{align}
\end{minipage}

\vspace{0.3cm}\noindent
As one can check, each field on the rhs in the condition \eqref{vecp2} has opposite signs compared with the corresponding one in \eqref{vecp1}.
The condition \eqref{vecp2} makes the Lagrangian $\mc{L}_{\text{YM}}$ invariant.
Let us explain why the doublet with opposite charges in the matter sector is necessary.
Now, the parity conditions for the differential operators are
\begin{align}
\pa_\compe \to - \pa_\compe,
\quad
\pa_{\compt, \compd} \to + \pa_{\compt,\compd}.
\end{align}
Accordingly, the covariant derivative behaves as
$( \pa_\mu - i \ch A_\mu^{(\mc{CP})}) \to \pm ( \pa_\mu + i \ch A_\mu^{(\mc{CP})})$ under \eqref{vecp2}.
It means that the parity condition \eqref{vecp2} works as \textit{charge conjugation}.
Therefore, matter theories with single gauge charge $\ch$ is NOT invariant under 
\eqref{vecp2}.
On the other hand, however, an interesting phenomenon happens if we turn on a \textit{doublet}  \eqref{ematp2} with opposite charges.
In this case, the following transformations of Lagrangians occur:
\vspace{-1cm}
\begin{align}
&
\qquad \qquad \qquad  \qquad \quad \quad \quad \  \
\begin{xy}
{(0,-10) \ar @/^3mm/ (15,-10)}
\end{xy}
\notag 
\\
&
\xymatrix{
\mc{L}\ar @(ul,dl)}
\hspace{-1mm}
_{\text{YM}}(V^{(\mc{CP})})+
\mc{L}_{\text{mat}} (\Phi_1 ; +\ch) +  \mc{L}_{\text{mat}} (\Phi_2 ; - \ch).
\label{YM_mat2}
\\
&
\qquad \qquad \qquad  \qquad \quad \quad \quad \ \
\begin{xy}
{(15,10) \ar @/^3mm/ (0,10)}
\end{xy}
\notag
\end{align}

\vspace{-0.3cm}\noindent
Compared with the previous one in \eqref{YM_mat},
the invariance displayed in \eqref{YM_mat2} looks exotic.
It means that the two matter multiplets merge into a \textit{doublet} and define a single matter on $\Sp^2 \times \Sp^1$.
As a result, we can recycle known calculations with slight modifications caused by an unorientable structure of $\Rp^2 \times \Sp^1$. We explain the details in Appendix \ref{LOCALIZATION}.

\paragraph{BF term}
As we noted,
the supersymmetric BF term is impossible within only $V^{(\mc{P})}$ or $V^{(\mc{CP})}$. However, a hybrid of $V^{(\mc{P})}$ and $V^{(\mc{CP})}$ 
gives a Lagrangian invariant under a set of the parity transformations on $\Rp^2 \times \Sp^1$ as follows:
\begin{align}
\mc{L}_{\text{BF}} (V^{(\mc{P})} , V^{(\mc{CP})})
=
A
^{(\mc{P})}
\wedge
  F
^{(\mc{CP})}
- \olam^{(\mc{P})} \lam^{(\mc{CP})}
- \olam^{(\mc{CP})} \lam^{(\mc{P})} 
+ 2 D^{(\mc{P})} \sigma^{(\mc{CP})}
+ 2 D^{(\mc{CP})} \sigma^{(\mc{P})} 
,
\label{BFt}
\end{align}
because the parity-odd feature of $V^{(\mc{P})}$ is totally cancelled by the opposite parity feature of $V^{(\mc{CP})}$.
Thus, we take this term into account seriously from now on.

\newpage
\subsection{Recipe of index for abelian gauge theories}\label{Recipe}
We summarize here the most general formulas for abelian gauge theories' indices on $\Rp^2 \times \Sp^1$
\begin{align}
\mathcal{I} ^{\mathbb{RP}^2}_{\text{Theory}} ( x, \alpha ) = \text{Tr}_{\mathcal{H}_{\text{Theory}} ^{\mathbb{RP}^2}} \Big( ( - 1 )^{\hat{F}} 
x^{\hat{H} + \hat{j}_{3}} \alpha^{\hat{f}} \Big),
\label{sci}
\end{align}
where $\mathcal{H}_{\text{Theory}} ^{\mathbb{RP}^2}$ represents the Hilbert space of the theory on $\Rp^2$, $\hat{F}$ is the fermion number, $\hat{H}$ is the energy which satisfies the BPS condition $\hat{H} + \hat{R} - \hat{j}_{3} = 0$, $\hat{R}$ is the $R$-symmetry, and $\hat{j}_3$ is the third component of the orbital angular momentum (quantities with hats are operators).
This quantity can be expressed by the path integral of the SUSY theories on $\Rp^2 \times \Sp^1$.
See the literatures \cite{Bhattacharya:2008zy, Kim:2009wb, Imamura:2011su, Tanaka:2014oda} for more details.
\paragraph{Quiver rules}
To make our recipe comprehensive, we define quiver rules.
We define nodes for two types of the vector multiplets defined in \eqref{vecp1} and \eqref{vecp2},
\begin{align}
&V^{(\mc{P})}
:
\xymatrix {
*++[o][F-]{\text{\scriptsize$\mc{P}$}} 
}
,
\qquad
V^{(\mc{CP})}
:
\xymatrix {
*++[o][F-]{\text{\tiny$\mc{CP}$}} 
}.
\end{align}
In addition, we introduce two types of matter multiplets defined in \eqref{ematp1} and \eqref{ematp2},
\begin{align}
&\Phi_s
:
\xymatrix{
*+[F-]{\Phi}
}
\ , \qquad
\Phi_d
:
\vspace{1cm}
\xymatrix{
*+[F-]{^{\Phi_1}_{\Phi_2}}
}\ ,
\end{align}
The gauge coupling is represented by an arrow 
connecting between a rectangle and a node.
The direction means the sign of the U$(1)$ gauge charge, e.g. 
$\xymatrix{
*+[F-]{\Phi }
\ar@{->}[r]
&
*++[o][F-]{} 
}
\Leftrightarrow \text{sign}(\ch)>0
$,
 and the number of arrowheads means the absolute value of the gauge charge.
For example, $\xymatrix{
\ar@{->>}[r]
&
}$
means $|\ch|=2$.
$\xymatrix{
\ar@{->}[r]_{\underbrace{}_{\ch}}
&
}$
means the generic $\ch$ instead of writing many arrowheads.

\paragraph{Possible couplings}
As already noted, the possible interactions on $\Rp^2 \times \Sp^1$ are restricted, and all of them are listed as follows:
\begin{align}
&\bullet \
\text{Matter - Matter}
:
\mc{W}( \Phi ),
\\
&\bullet \
\text{Vector - Vector}:
\
\xymatrix{
*++[o][F-]{\text{\scriptsize$\mc{P}$}} 
\ar @{-}[r]^{\text{BF}}
&
*++[o][F-]{\text{\tiny$\mc{CP}$}} 
},
\\
&\bullet \
\text{Matter - Vector}:
\
\xymatrix{
*+[F-]{\Phi}
\ar@{->}[r]^<{}_{\underbrace{}_{\ch}} 
&
*++[o][F-]{\text{\scriptsize$\mc{P}$}} 
},
\quad
\xymatrix{
*+[F-]{^{\Phi_1}_{\Phi_2}}
\ar @<1mm> [r]
\ar @<-1mm> [r]_{\underbrace{}_{\ch}}
&
*++[o][F-]{\text{\scriptsize$\mc{P}$}} 
},
\quad
\xymatrix{
*+[F-]{^{\Phi_1}_{\Phi_2}}
\ar @<1mm> [r]
&
*++[o][F-]{\text{\tiny$\mc{CP}$}} 
\ar @<1mm> [l]^{\underbrace{}_{\ch}}
},
\end{align}
where we do not define quiver rules for a superpotential $\mathcal{W}$
because the contribution of $\mathcal{W}$ in \textit{Coulomb branch localization} becomes irrelevant\footnote{
The allowed superpotentials are of course the ones satisfying parity invariance.} eventually.
One can draw an arbitrary quiver by using these rules to define meaningful Abelian SUSY theories on $\Rp^2 \times \Sp^1$.
\paragraph{Formulas for the index \eqref{sci}}
We can get the formulas for the exact indices based on the localization techniques.
For the details, see \cite{Tanaka:2014oda} and Appendix \ref{LOCALIZATION}. 
Here, we just summarize the recipe for them.
First of all, we replace each node as follows.
\begin{align}
&\xymatrix {
*++[o][F-]{\text{\scriptsize$\mc{P}$}} 
}
=
x^{+ \frac{1}{4}} \frac{(x^4 ; x^4)_\infty}{(x^2 ; x^4)_\infty}
\sum_{B^{\pm}_{}}
\frac{1}{2 \pi}
\int_0^{2\pi} d \theta \
\xymatrix {
*++[o][F.]{\text{\scriptsize$B^\pm, \theta$}} 
}
,
\label{Psum}
\\
&\xymatrix {
*++[o][F-]{\text{\tiny$\mc{CP}$}} 
}
=
x^{- \frac{1}{4}} \frac{(x^2 ; x^4)_\infty}{(x^4 ; x^4)_\infty}
\sum_{B \in 2 \mathbb{Z}}
\frac{1}{2}
\sum_{\theta_\pm} 
\xymatrix {
*++[o][F.]{\text{\scriptsize$B , \theta_\pm$}} 
}
,
\label{CPsum}
\end{align}
where the dotted node $
\xymatrix {
*++[o][F.]{} 
}
$
means background vector multiplets given by
\begin{align}
\xymatrix {
*++[o][F.]{\text{\scriptsize$B^\pm, \theta$}}
}
&:
A=B_{}^\pm A_{\text{flat}} + \frac{\theta}{2 \pi} d\compd , 
\quad
\si =0,
\qquad \ \
\Big(
\left. \begin{array}{ll}
B^+=0  \\
B^- = 1  
\end{array} \right.
\quad
\theta \in[0, 2\pi]
\Big)
\label{PBG}
,
\\
\xymatrix {
*++[o][F.]{\text{\scriptsize$B,\theta_\pm$}} 
}
&:
A=B \ A_{\text{mon}} + \frac{\theta_\pm}{2 \pi} d \compd , 
\quad
\si=- \frac{B}{2} , 
\ \ \ \ 
\Big(
B \in 2 \mathbb{Z},
\quad
\left. \begin{array}{ll}
\theta_+=0  \\
\theta_- =\pi
\end{array} \right.
\Big)
\label{CPBG}
,
\end{align}
where $A_{\text{flat}}$ is the nontrivial flat connection\footnote{In the language used in \cite{Tanaka:2014oda}, $A_{\text{flat}} = A_{\text{flat}}^-$.} on $\Rp^2$, $A_{\text{mon}}$ is the Dirac monopole configuration with monopole charge $1$, and $\theta$ is a Wilson line phase along $\mathbb{S}^{1}$.
As the second step, we replace the rigid line with the dotted line:
$
\xymatrix{
\ar@{->}[r]
&
}
=
\xymatrix{
\ar@{.>}[r]
&
}.
$
Then, we get the integrand or the summand in \eqref{Psum} or \eqref{CPsum} with the following contributions\footnote{We correct the formula \eqref{d1loop} in the previous version by adding the exponential prefactor.} from the quiver diagrams including the matter multiplets\footnote{The matter multiplet is fixed with R-charge $\hat{R} = - \Delta$. In \eqref{d1loop} and \eqref{s1loop}, we can couple arbitrary number of nodes. If so, the contribution becomes the one by added appropriate terms in $\ch \theta$ and $\ch B$.}:
\begin{align}
\xymatrix{
*++[o][F.]{\text{\scriptsize$B^\pm_1 , \theta_1$}} 
&
*+[F-]{\Phi}
\ar@{.>}[l]^{\underbrace{}_{\ch_1}} 
\ar@{.>}[r]^{}_{\underbrace{}_{\ch_2}} 
&
*++[o][F.]{\text{\scriptsize$B^\pm_2 , \theta_2$}} 
}
&=
\left\{ \begin{array}{ll}
\Big(
x^{ \frac{\Delta - 1}{4}} e^{ \frac{i (\ch_1 \theta_1+\ch_2 \theta_2)}{4} } 
\Big)^{+1}
	\frac{( e^{- i (\ch_1 \theta_1 + \ch_2 \theta_2)} 
	x^{2 - \Delta} ; x^{4} )_{\infty}}{( e^{+ i (\ch_1 \theta_1 + \ch_2 \theta_2)} 
	 x^{0+\Delta} ; x^{4} )_{\infty}} 
	& \text{for} \  (-1)^{\ch_1 B^\pm_1 + \ch_2 B^\pm_2} = +1, \\[.5em]
\Big(
x^{ \frac{\Delta - 1}{4}} e^{ \frac{i (\ch_1 \theta_1 + \ch_2 \theta_2)}{4} } 
\Big)^{-1}
	\frac{( e^{- i (\ch_1 \theta_2 + \ch_2 \theta_2)} 
	x^{4 - \Delta} ; x^{4} )_{\infty}}{( e^{+ i (\ch_1 \theta_1 + \ch_2 \theta_2)} 
	 x^{2+\Delta} ; x^{4} )_{\infty}} & 
	\text{for}\ (-1)^{\ch_1 B^\pm_1 + \ch_2 B^\pm_2} = -1, \\
\end{array} \right.
\label{s1loop}
\\
\xymatrix{
*++[o][F.]{\text{\scriptsize$B^\pm_1 , \theta_1$}} 
\ar@{<.}[r]<1mm>
\ar@{<.}[r]<-1mm>_{\underbrace{}_{\ch_1}}
&
*+[F-]{^{\Phi_1}_{\Phi_2}}
\ar@{.>}[r]<1mm>
\ar@{<.}[r]<-1mm>_{\underbrace{}_{\ch_2}}
&
*++[o][F.]{\text{\scriptsize$B_2 , \theta{_2}{_\pm}$}} 
}
&=
e^{i \frac{\ch_1 \ch_2 B_{1}^{\pm} \theta{_2}{_\pm}}{2}}
 \Big(  x^{1-\Delta} e^{- i \ch_1 \theta_1} 
\Big)^{\frac{|\ch_2 B_2|}{2}}
\frac{\Big(
e^{-i ( \ch_1 \theta{_1} +\ch_2 \theta{_2}{_\pm}) }  
x^{|\ch_2 B_2|+(2-\Delta)} ; x^2 \Big)_\infty}
{\Big(
e^{+i (\ch_1 \theta{_1}+\ch_2 \theta{_2}{_\pm}  )}  
x^{|\ch_2 B_2|+(0+\Delta)} ; x^2\Big)_\infty},
\label{d1loop}
\\
\xymatrix{
*++[o][F.]{\text{\scriptsize$B^\pm, \theta$}}
\ar @{.}[r]^{\text{BF}}
&
*++[o][F.]{\text{\scriptsize$B, \theta_\pm$}}
}&=
(-1)^{B^\pm \theta_\pm /\pi  }
\times
\Big( e^{i \theta}
\Big)
^{B/2}
.
\label{BFvalue}
\end{align}
\paragraph{Weakly gauged global symmetry}
If the system has a global symmetry, we can turn on a background vector multiplet $\mc{V}$ which couples to the global symmetry current.
We can take its effect into account by turning on the corresponding dotted node \eqref{PBG} or \eqref{CPBG} from the beginning.
For example, a typical diagram is as follows:
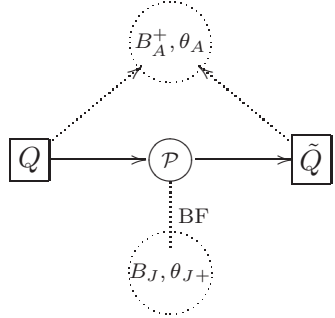
\begin{align}
\xymatrix {
&
*++[o][F.]{\text{\scriptsize$B^+_A, \theta_A$}} 
&
\\
*+[F-]{\text{$Q$}} 
\ar@{->}[r]
\ar@{.>}[ur]
&
*++[o][F-]{\text{\scriptsize$\mc{P}$}} 
\ar@{->}[r]
\ar@{.}[d]^{\text{BF}}
&
*+[F-]{\text{$\tilde{Q}$}}
\ar@{.>}[ul]
\\
&
*++[o][F.]{\text{\scriptsize$B_J,\theta{_J}{_+}$}} 
&
}	\notag
\end{align}
It means that there are two background gauge fields $\mathcal{V}_{A}$ and $\mathcal{V}_{J}$ coupled to $Q,\tilde{Q}$, and $V^{(\mc{P})}$, respectively.
We can easily derive the corresponding index 
by repeating the above procedure 
without the summation or the integration in \eqref{Psum} and \eqref{CPsum} for $B_A^+,\theta_A$ and $B_J,\theta{_J}{_+}$.
Note that the appropriate parity type of the background field should be selected from a flavor charge of each field under the corresponding global symmetry.


\bibliographystyle{common_supp_files/sonota/utphys}
\bibliography{common_supp_files/Ref}


\section{Mirror symmetry on $\mathbb{RP}^{2} \times \mathbb{S}^{1}$}\label{Mirrorsym}
\begin{table}[t]
\vspace{-1em}
\begin{center}
\begin{tabular}{c}
	\begin{minipage}[b]{.48\hsize} 
		\begin{center}
			\begin{tabular}{|c|c|c|c|c|} \hline
			& U$(1)$ & U$(1)_{J}$ & U$(1)_{A}$ & $\hat{R}$ \\ \hline \hline
			$Q$ & $+ 1$ & $0$ & $+ 1$ & $- \Delta$ \\ \hline
			$\tilde{Q}$ & $- 1$ & $0$ & $+ 1$ & $- \Delta$ \\ \hline
			\end{tabular}
		\end{center}
	\vspace{-.75em}
	\caption{Charges in the SQED. Global symmetries are a topological U$(1)_{J}$ and an axial U$(1)_{A}$.\label{cSQED}}
	\end{minipage}
	\hspace{2em}
	\begin{minipage}[b]{.52\hsize} 
		\begin{center}
			\begin{tabular}{|c|c|c|c|} \hline
			& U$(1)_{V}$ & U$(1)_{A}$ &$\hat{R}$ \\ \hline \hline
			$X$ & $+ 1$ & $+ 1$ & $- ( 1 - \Delta )$ \\ \hline
			$Y$ & $- 1$ & $+ 1$ & $- ( 1 - \Delta )$ \\ \hline
			$Z$ & 0 & $- 2$ & $- 2 \Delta$ \\ \hline
			\end{tabular}
		\end{center}
	\vspace{-.75em}
	\caption{Charges in the XYZ model. Global symmetries are a vector-like U$(1)_{V}$ and an axial U$(1)_{A}$.\label{cXYZ}}
	\end{minipage}
\end{tabular}
\end{center}
\end{table}

\begin{table}[t] 
\vspace{-.7em}
\begin{center}
	\begin{tabular}{ccc} \hline
	SQED & & XYZ \\ \hline
	U$(1)_{J}$, $J_{J}$ & $\leftrightarrow$ & U$(1)_{V}$, $J_{V}$ \\
	U$(1)_{A}$, $J_{A}$ & $\leftrightarrow$ & U$(1)_{A}$, $- J_{A}$ \\
	$e^{\left( \sigma + i \rho \right) / e^{2}}$, $e^{- \left( \sigma + i \rho \right) / e^{2}}$ & $\leftrightarrow$ & $X$, $Y$ \\
	$Q \tilde{Q}$ & $\leftrightarrow$ & $Z$ \\ \hline
	\end{tabular}
\end{center}
\vspace{-1em}
\caption{The mirror map. $J_{J}$, $J_{V}$, and $J_{A}$ are currents associated to U$(1)_{J}$, U$(1)_{V}$, and U$(1)_{A}$, respectively. $e^{2}$ is a coupling constant.\label{mmap}}
\end{table}

Mirror symmetry is a kind of dualities between theories which flow to the same IR fixed point along the RG flow. In this paper, we deal with $\mathcal{N} = 2$ mirror symmetry \cite{Intriligator:1996ex, Aharony:1997bx} between the SQED and the XYZ model on $\mathbb{RP}^{2} \times \mathbb{S}^{1}$; the former has a vector multiplet and two chiral fields $Q$ and $\tilde{Q}$ charged under the U$(1)$ gauge group, and the latter is comprised of three chiral fields $X, Y$, and $Z$ coupled to each other through the superpotential $\mathcal{W} = XYZ$. In addition, they contain two global U$(1)$ symmetries and $R$-symmetry acting on the chiral fields. We arrange all charge assignments in Table \ref{cSQED} and \ref{cXYZ}.

In the context of mirror symmetry, the languages of the SQED are mapped to those of the XYZ model as in Table \ref{mmap}. The bottom two lines in Table \ref{mmap} are the correspondences of moduli parameters. $\pm ( \sigma + i \rho )$, where $\rho$ is the dual photon defined by
\begin{align} 
\frac{1}{2} \epsilon_{\mu \nu \rho} F^{\nu \rho} = \partial_{\mu} \rho,
\label{dualphoton}
\end{align}
characterize the Coulomb branch, and the Higgs branch is parametrized by $Q \tilde{Q}$. Notice that we assume the maps of the moduli are kept not only at the classical level but also for quantum fluctuations of these fields. We shall follow this assumption to decide the parity condition for the XYZ model.

We consider the case where global symmetries are gauged. 
The fact that the background gauge field of a topological U$(1)_{J}$ in the SQED is coupled to a topological current $J_{T} = \ast F$ shows the appearance of a BF term as a classical contribution. In fact, our BF term \eqref{BFvalue} carries out a significant effect to manifest mirror symmetry. With the gauging prescriptions as explained above, we find four types of mirror symmetry depending on the choices of consistent parity conditions in the SQED and the XYZ model. The first one explained in the subsection \ref{Ptype} is an one-parameter extension of mirror symmetry demonstrated in \cite{Tanaka:2014oda}, and others are completely new. We give the readers proof for all types of $\mathcal{N} = 2$ mirror symmetry in Apeendix \ref{24mirrorsym}.

\subsection{The map of parity conditions}
\vspace{-.5em}
\begin{table}[h] 
\begin{tabular}{c}
	\hspace{2em}
	\begin{minipage}[t]{.5\hsize} 
		\begin{align*}
			\xymatrix {
                         *+[F-]{\text{$Q$}} 
                         \ar@{->}[r]
                         &
                         *++[o][F-]{\text{\scriptsize$\mc{P}$}} 
                         \ar@{->}[r]
                         &
                         *+[F-]{\text{$\tilde{Q}$}} }
		\end{align*}
		\vspace{-.3em}
		\begin{center}
			\begin{tabular}{rcl} \hline
			\multicolumn{3}{c}{SQED$^{( \mathcal{P} )}$} \\ \cline{1-3}
			$\sigma, \rho$ & $\begin{xy} {(0,5) \ar (7.5,5)} \end{xy}$ & $- \sigma, - \rho$  \\
			$Q$ & $\begin{xy} {(0,5) \ar (7.5,5)} \end{xy}$ & $Q$ \\
			$\tilde{Q}$ & $\begin{xy} {(0,5) \ar (7.5,5)} \end{xy}$ & $\tilde{Q}$ \\ \hline
			\end{tabular}
		\end{center}
	\end{minipage}
	\hspace{-3.5em}
	$\begin{xy} {(0,-10) \ar@{<=>} (10,-10)} \end{xy}$
	\hspace{-5em}
	\begin{minipage}[t]{.5\hsize} 
		\begin{align*}
			\hspace{-3em}
			\xymatrix {
                 	&
                 	*+[F-]{\text{\scriptsize$
                 	\begin{matrix}
                 	X &
                 	Y
                 	\end{matrix}
                 	$}} 
                 	&
                 	*+[F-]{\text{$Z$}} }
		\end{align*}
		\vspace{-.3em}
		\begin{center}
			\begin{tabular}{rcl} \hline
			\multicolumn{3}{c}{$^{\text{X}}_{\text{Y}} \text{Z}$} \\ \cline{1-3}
			\shortstack{\strut $X$ \\ $Y$ } & $\begin{xy} {(0,10) \ar (7.5,5)}, {(0,5) \ar (7.5,10)} \end{xy}$ & \shortstack{\strut $Y$ \\ $X$ } \\
			$Z$ & $\begin{xy} {(0,5) \ar (7.5,5)} \end{xy}$ & $Z$ \\ \hline
			\end{tabular}
		\end{center}
	\end{minipage}
\end{tabular}
\caption{Parity conditions in SQED$^{( \mathcal{P} )}$ and $^{\text{X}}_{\text{Y}} \text{Z}$.\label{b1-type}}
\end{table}

In this subsection, we summarize the parity actions in the SQED and the corresponding conditions in the XYZ model based on the mirror map of the moduli parameters.

Firstly, we focus on the SQED under the parity conditions \eqref{vecp1} and \eqref{ematp1} denoted by SEQD$^{( \mathcal{P} )}$. In the language of the moduli parameters, $\sigma$ and $\rho$ receive flipping its sign under $\mathcal{P}$, but $Q$ and $\tilde{Q}$ do not mix up. Recalling the mirror map (Table \ref{mmap}), we can see that changing the signs of $\sigma$ and $\rho$ corresponds to the interchange of $X$ and $Y$, and $Z$ does not take any effect (the right side of Table \ref{b1-type}). This means that $( X, Y )$ behave as a doublet matter multiplet described by the parity condition \eqref{ematp2}. We name this dual theory $^{\text{X}}_{\text{Y}} \text{Z}$ to represent combining $( X, Y )$ as a single multiplet on $\mathbb{S}^{2} \times \mathbb{S}^{1}$. We argue mirror symmetry between SEQD$^{( \mathcal{P} )}$ and $^{\text{X}}_{\text{Y}} \text{Z}$ with respect to the parity conditions in Table \ref{b1-type}.

\begin{table}[h] 
\begin{tabular}{c}
	\hspace{2em}
	\begin{minipage}[t]{.5\hsize} 
		\begin{align*}
		\xymatrix {
			*+[F-]{\text{\scriptsize$
			\begin{matrix}
			Q \\
			\tilde{Q}
			\end{matrix}
			$}}
			&
			*++[o][F-]{\text{\scriptsize$\mc{CP}$}} 
			\ar@{->}[l]<1.5mm>
			\ar@{<-}[l]<-1.5mm> }
		\end{align*}
		\vspace{-.3em}
		\begin{center}
			\begin{tabular}{rcl} \hline
			\multicolumn{3}{c}{SQED$^{( \mathcal{CP} )}$} \\ \cline{1-3}
			$\sigma, \rho$ & $\begin{xy} {(0,5) \ar (7.5,5)} \end{xy}$ & $+ \sigma, + \rho$  \\
			\shortstack{ $Q$ \\ $\tilde{Q}$ } & $\begin{xy} {(0,10) \ar (7.5,5)}, {(0,5) \ar (7.5,10)} \end{xy}$ & \shortstack{\strut $\tilde{Q}$ \\ $Q$ } \\ \hline
			\end{tabular}
		\end{center}
	\end{minipage}
	\hspace{-4em}
	$\begin{xy} {(0,-10) \ar@{<=>} (10,-10)} \end{xy}$
	\hspace{-4.5em}
	\begin{minipage}[t]{.5\hsize} 
		\begin{align*}
			\xymatrix {
			*+[F-]{\text{$X$}} 
			&
			*+[F-]{\text{$Y$}} 
			&
			*+[F-]{\text{$Z$}} }
		\end{align*}
		\vspace{-.15em}
		\begin{center}
			\begin{tabular}{ccc} \hline
			\multicolumn{3}{c}{XYZ} \\ \cline{1-3}
			$X$ & $\begin{xy} {(0,5) \ar (7.5,5)} \end{xy}$ & $X$ \\
			$Y$ & $\begin{xy} {(0,5) \ar (7.5,5)} \end{xy}$ & $Y$ \\
			$Z$ & $\begin{xy} {(0,5) \ar (7.5,5)} \end{xy}$ & $Z$ \\ \hline
			\end{tabular}
		\end{center}
	\end{minipage}
\end{tabular}
\caption{Parity conditions in SQED$^{( \mathcal{CP} )}$ and XYZ.\label{b3-type}}
\end{table}
Next, we consider the SQED satisfying our new parity conditions \eqref{vecp2} and \eqref{ematp2} which we call SQED$^{( \mathcal{CP} )}$. The moduli $\sigma$ and $\rho$ transform trivially, while the matters $Q$ and $\tilde{Q}$ must be exchanged each other because of accompanying charge conjugation $\mathcal{C}$ with $\mathcal{P}$ (the left side of Table \ref{b3-type}). The mirror map of the moduli spaces (Table \ref{mmap}) gives us the parity conditions in the XYZ model (denoted just by XYZ) such that $X$ and $Y$ remain intact, and also $Z$ is not affected since it is the product of $Q$ and $\tilde{Q}$. This correspondence of the parity conditions is indicated in Table \ref{b3-type} of which we make use in verifying mirror symmetry between SQED$^{( \mathcal{CP} )}$ and XYZ.

\subsection{$\text{SQED}^{( \mathcal{P} )}$ vs $^{\text{X}}_{\text{Y}} \text{Z}$}
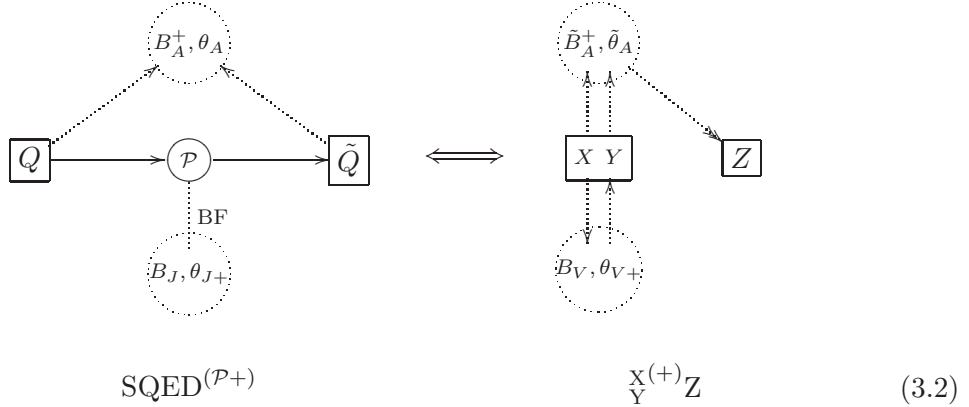
\label{Ptype}
\begin{align}
\xymatrix {
&
*++[o][F.]{\text{\scriptsize$B_{A}^{+}, \theta_A$}} 
&
\\
*+[F-]{\text{$Q$}} 
\ar@{->}[r]
\ar@{.>}[ur]
&
*++[o][F-]{\text{\scriptsize$\mc{P}$}} 
\ar@{->}[r]
\ar@{.}[d]^{\text{BF}}
&
*+[F-]{\text{$\tilde{Q}$}}
\ar@{.>}[ul]
\\
&
*++[o][F.]{\text{\scriptsize$B_J, \theta_{J +}$}} 
&
\\
&
\text{SQED}^{(\mc{P}+)}
&
}	
\ \ \ \quad
\begin{xy}
{(5,-15) \ar @{<=>} (15,-15)}
\end{xy}
\hspace{-1.5em}
\xymatrix {
&
*++[o][F.]{\text{\scriptsize$\tilde{B}_{A}^{+}, \tilde{\theta}_A$}} 
\ar@{<.}[d]<1.5mm>
\ar@{<.}[d]<-1.5mm>[r]
&
\\
&
*+[F-]{\text{\scriptsize$
\begin{matrix}
X &
Y
\end{matrix}
$}} 
\ar@{<.}[d]<1.5mm>
\ar@{.>}[d]<-1.5mm>
&
*+[F-]{\text{$Z$}}
\ar@{<<.}[ul]
\\
&
*++[o][F.]{\text{\scriptsize$B_V, \theta_{V +}$}} 
&
}
\notag \\[-1.95em]
&\hspace{-4.5em}
{^\text{X}
_\text{Y}}^{(+)}
\text{Z}
\label{quiverP+} \\[-1em]
\notag
\end{align}

\paragraph{SQED$^{( \mathcal{P} + )}$ vs $^{\text{X}}_{\text{Y}}{}^{( + )}\text{Z}$}
This is the case studied in \cite{Tanaka:2014oda} 
without the BF term in the SQED. Here, we incorporate the BF term into the index by turning on the $\mc{CP}$-type background gauge field \eqref{CPBG} for U$(1)_J$ global symmetry parametrized by $(B_J, \theta_{J+})$.
$\mc{CP}$-type is necessary to make the BF term parity-even. 
In addition, we turn on the $\mc{P}$-type background gauge field \eqref{PBG} for U$(1)_A$ global symmetry parametrized by $(B_A^+,\theta_A)$.
$\mc{P}$-type is necessary to make the gauge coupling parity-even. 
On the left side of \eqref{quiverP+}, we draw the quiver diagram.
We call it SQED$^{(\mc{P}+)}$ where the superscript indicates the existence of $\theta_{J+}$. 
The index is given by using the quiver rules summarized in subsection \ref{Recipe} as follows.
\begin{align} 
&
\mathcal{I}_{\text{SQED}^{( \mathcal{P} + )}} ( x, \alpha, B_{J} ) \notag \\ 
&=
q^{\frac{1}{8}}
\frac{( q^{2} ; q^{2} )_{\infty}}{( q ; q^{2} )_{\infty}}
\oint_{C_0} \frac{d z}{2 \pi i z} z^{s}
\left\{
a^{- \frac{1}{4}}
\frac
{( z^{- 1} a^{\frac{1}{2}} q^{\frac{1}{2}}, z a^{\frac{1}{2}} q^{\frac{1}{2}}; q^{2} )_{\infty}}
{( z a^{- \frac{1}{2}} q^{\frac{1}{2}}, z^{- 1} a^{- \frac{1}{2}} q^{\frac{1}{2}} ; q^{2} )_{\infty}}
+
a^{\frac{1}{4}}
\frac
{( z^{- 1} a^{\frac{1}{2}} q^{\frac{3}{2}}, z a^{\frac{1}{2}} q^{\frac{3}{2}} ; q^{2} )_{\infty}}
{( z a^{- \frac{1}{2}} q^{\frac{3}{2}}, z^{- 1} a^{- \frac{1}{2}} q^{\frac{3}{2}} ; q^{2} )_{\infty}} \right\}, 
\label{scib1sqed1}
\end{align}
where the integration contour $C_{0}$ is the unit circle, $z := e^{i \theta}$, and $s := B_{J}/2 \in \mathbb{Z}$ is a backgorund monopole flux. $\alpha = e^{i \theta_{A}}$ represents a fugacity associated to U$(1)_{A}$, and also we define $q = x^{2}$ and $a = \alpha^{- 2} x^{2 ( 1 - \Delta )}$ for latter use.

On the other hand, we write down the quiver diagram of the dual theory ${^{\text{X}}_{\text{Y}}}^{(+)}\text{Z}$ in \eqref{quiverP+} where again 
the superscript means the existence of $\theta_{V+}$.
We also turn on $B_V$ and $(\tilde{B}_A^{+},\tilde{\theta}_A)$.
Then, the quiver rules give the corresponding index as
\begin{align} 
\mathcal{I}_{{^{\text{X}}_{\text{Y}}}^{(+)}\text{Z}} ( x, \tilde{\alpha}, B_{V} )
=
q^{\frac{1}{8}} \tilde{a}^{- \frac{1}{4}}
\left( \tilde{a}^{- \frac{1}{2}} q^{\frac{1}{2}} \right)^{| \tilde{s} |}
\frac{( \tilde{a}^{- \frac{1}{2}} q^{1 + |\tilde{s}|} ; q )_{\infty}}{( \tilde{a}^{\frac{1}{2}} q^{|\tilde{s}|} ; q )_{\infty}}
\frac{( \tilde{a} ; q^{2} )_{\infty}}{( \tilde{a}^{- 1} q ; q^{2} )_{\infty}},
\label{scib1xyz1}
\end{align}
where we define $\tilde{\alpha} = e^{i \tilde{\theta}_A}$, $\tilde{a} = \tilde{\alpha}^{2} x^{2 ( 1 - \Delta )}$ and $\tilde{s} = B_{V}/2 \in \mathbb{Z}$.

According to $\mc{N}=2$ mirror symmetry, 
the following identity is expected to hold:
\begin{align}
\mathcal{I}_{\text{SQED}^{( \mathcal{P} + )}} ( x, \alpha, B_{J} )
=
\mathcal{I}_{{^{\text{X}}_{\text{Y}}}^{(+)}\text{Z}} ( x, \tilde{\alpha}, B_{V} ),
\label{mirror1}
\end{align}
under identifications $\alpha = \tilde{\alpha}^{- 1}$ and $B_{J} = B_{V}$ found from the mirror map (Table \ref{mmap}). Note that, for all cases in the paper, we fix the background gauge field for U$(1)_{A}$ to be $B_{A}^{+}$.

\begin{align}
\xymatrix {
&
*++[o][F.]{\text{\scriptsize$B_{A}^{+}, \theta_A$}} 
&
\\
*+[F-]{\text{$Q$}} 
\ar@{->}[r]
\ar@{.>}[ur]
&
*++[o][F-]{\text{\scriptsize$\mc{P}$}} 
\ar@{->}[r]
\ar@{.}[d]^{\text{BF}}
&
*+[F-]{\text{$\tilde{Q}$}}
\ar@{.>}[ul]
\\
&
*++[o][F.]{\text{\scriptsize$B_J, \theta_{J -}$}} 
&
\\
&
\text{SQED}^{(\mc{P}-)}
&
}
\ \ \ \quad
\begin{xy}
{(5,-15) \ar @{<=>} (15,-15)}
\end{xy}
\hspace{-1.5em}
\xymatrix {
&
*++[o][F.]{\text{\scriptsize$\tilde{B}_{A}^{+}, \tilde{\theta}_A$}} 
\ar@{<.}[d]<1.5mm>
\ar@{<.}[d]<-1.5mm>[r]
&
\\
&
*+[F-]{\text{\scriptsize$
\begin{matrix}
X &
Y
\end{matrix}
$}} 
\ar@{<.}[d]<1.5mm>
\ar@{.>}[d]<-1.5mm>
&
*+[F-]{\text{$Z$}}
\ar@{<<.}[ul]
\\
&
*++[o][F.]{\text{\scriptsize$B_V, \theta_{V -}$}}
&
}
\notag \\[-1.95em]
&\hspace{-4.5em}
{^\text{X}
_\text{Y}}^{(-)}
\text{Z}
\label{quiverP-} \\[-.5em]
\notag
\end{align}

\paragraph{SQED$^{( \mathcal{P} - )}$ vs ${^{\text{X}}_{\text{Y}}}^{(-)}\text{Z}$}
We can define the SQED with the same parity conditions as used in SQED$^{( \mathcal{P} + )}$ but choosing the other background Wilson line phase, that is, $\theta_{J -}$. We name this case SQED$^{( \mathcal{P} - )}$ whose quiver diagram is drawn on the left side of \eqref{quiverP-}. This quiver diagram instantly gives us the index of SQED$^{( \mathcal{P} - )}$
\begin{align} 
&
\mathcal{I}_{\text{SQED}^{( \mathcal{P} - )}} ( x, \alpha, B_{J} ) \notag \\ 
&=
q^{\frac{1}{8}} \frac{( q^{2} ; q^{2} )_{\infty}}{( q ; q^{2} )_{\infty}}
\oint_{C_0} \frac{d z}{2 \pi i z} z^{s}
\left\{
a^{- \frac{1}{4}}
\frac
{( z^{- 1} a^{\frac{1}{2}} q^{\frac{1}{2}}, z a^{\frac{1}{2}} q^{\frac{1}{2}}; q^{2} )_{\infty}}
{( z a^{- \frac{1}{2}} q^{\frac{1}{2}}, z^{- 1} a^{- \frac{1}{2}} q^{\frac{1}{2}}; q^{2} )_{\infty}}
-
a^{\frac{1}{4}}
\frac
{( z^{- 1} a^{\frac{1}{2}} q^{\frac{3}{2}}, z a^{\frac{1}{2}} q^{\frac{3}{2}} ; q^{2} )_{\infty}}
{( z a^{- \frac{1}{2}} q^{\frac{3}{2}}, z^{- 1} a^{- \frac{1}{2}} q^{\frac{3}{2}} ; q^{2} )_{\infty}}
\right\}, 
\label{scib2sqed1}
\end{align}
where all variables are defined in the same manner as $\mathcal{I}_{\text{SQED}^{( \mathcal{P} + )}}$ \eqref{scib1sqed1}. The only difference from the previous example is the relative weight originated from the BF term \eqref{BFvalue} in summing over the dynamical gauge holonomies in \eqref{CPsum}. Thus, we can proceed with the computation in a parallel way to \eqref{scib1sqed1}.

On the dual side, we call it ${^{\text{X}}_{\text{Y}}}^{(-)}\text{Z}$, although chiral fields satisfy the same parity conditions as ${^{\text{X}}_{\text{Y}}}^{(+)}\text{Z}$, they couple with the background Wilson line phase $\theta_{V -}$. 
The index of ${^{\text{X}}_{\text{Y}}}^{(-)}\text{Z}$ corresponding to the quiver on the right side of \eqref{quiverP-} is given by
\begin{align} 
\mathcal{I}_{{^{\text{X}}_{\text{Y}}}^{(-)}\text{Z}} ( x, \tilde{\alpha}, B_{V} )
=
q^{\frac{1}{8}} \tilde{a}^{- \frac{1}{4}}
\left( \tilde{a}^{- \frac{1}{2}} q^{\frac{1}{2}} \right)^{| \tilde{s} |}
\frac{( - \tilde{a}^{- \frac{1}{2}} q^{1 + |\tilde{s}|} ; q )_{\infty}}{( - \tilde{a}^{\frac{1}{2}} q^{|\tilde{s}|} ; q )_{\infty}}
\frac{( \tilde{a} ; q^{2} )_{\infty}}{( \tilde{a}^{- 1} q ; q^{2} )_{\infty}},
\label{scib2xyz1}
\end{align}
where we reuse the arguments $\tilde{a}$ and $\tilde{s}$ from \eqref{scib1xyz1}. The negative signs in the second fraction of \eqref{scib2xyz1} are caused by the fact that $X$, $Y$ which have U$(1)_{V}$ charges $\pm 1$ are coupled to $\theta_{V -}$.

Mirror symmetry between SQED$^{( \mathcal{P} - )}$ and ${^{\text{X}}_{\text{Y}}}^{(-)}\text{Z}$ should be realized as
\begin{align}
\mathcal{I}_{\text{SQED}^{( \mathcal{P} - )}} ( x, \alpha, B_{J} )
=
\mathcal{I}_{{^{\text{X}}_{\text{Y}}}^{(-)}\text{Z}} ( x, \tilde{\alpha}, B_{V} )
\label{mirror2}
\end{align}
under identifications $\alpha = \tilde{\alpha}^{- 1}$ and $B_{J} = B_{V}$ as before. The readers can confirm numerically two types of equivalence \eqref{mirror1} and \eqref{mirror2}. Also, they can be verified analytically by utilizing the $q$-binomial theorem \eqref{qbinomial}. We summarize mathematical tools and indicate exact proof for both examples in Appendix \ref{24mirrorsym} (see \cite{Tanaka:2014oda} for details and general arguments about the $q$-binomial theorem).


\clearpage
\subsection{$\text{SQED}^{( \mathcal{CP} )}$ vs XYZ}\label{CPtype}
\begin{align}
\xymatrix {
&
*++[o][F.]{\text{\scriptsize$B_{A}^{+}, \theta_A$}} 
&
\\
*+[F-]{\text{\scriptsize$
\begin{matrix}
Q \\
\tilde{Q}
\end{matrix}
$}}
\ar@{.>}[ur]<1.5mm>
\ar@{.>}[ur]<-1.5mm>
&
*++[o][F-]{\text{\scriptsize$\mc{CP}$}} 
\ar@{->}[l]<1.5mm>
\ar@{<-}[l]<-1.5mm>
\ar@{.}[d]^{\text{BF}}
&
\\
&
*++[o][F.]{\text{\scriptsize$B_{J}^{+},\theta_J$}} 
&
\\
&
\text{SQED}^{(\mc{CP}+)}
&
}
\hspace{-2em}
\begin{xy}
{(5,-15) \ar @{<=>} (15,-15)}
\end{xy}
\hspace{2em}
\xymatrix {
&
*++[o][F.]{\text{\scriptsize$\tilde{B}_{A}^{+}, \tilde{\theta}_A$}} 
\ar@{<.}[d]
\ar@{<.}[dl]
&
\\
*+[F-]{\text{$X$}} 
\ar@{.>}[dr]
&
*+[F-]{\text{$Y$}} 
\ar@{<.}[d]
&
*+[F-]{\text{$Z$}}
\ar@{<<.}[ul]
\\
&
*++[o][F.]{\text{\scriptsize$B_{V}^{+},\theta_V$}} 
&
}
\notag \\[-1.75em]
&\hspace{-7.5em}
\text{X}^{(+)}
\text{Y}^{(+)}
\text{Z}
\label{quiverCP+} \\[-1em]
\notag
\end{align}

\paragraph{SQED$^{( \mathcal{CP} + )}$ vs X$^{( + )}$Y$^{( + )}$Z}
Let us turn to concentrating on mirror symmetry under our new parity conditions \eqref{vecp2} and \eqref{ematp2} which act as charge conjugation $\mathcal{C}$ simultaneously with $\mathcal{P}$. Here, the background gauge field for U$(1)_{J}$ is in $\mc{P}$-type \eqref{PBG} to keep the BF term parity-even.
First, we choose it to be even-holonomy $B_{J}^{+}$ on the SQED side which we denote by SQED$^{( \mathcal{CP} + )}$. The quiver diagram of SQED$^{( \mathcal{CP} + )}$ is shown on the left side of \eqref{quiverCP+} where ($Q, \tilde{Q}$) should be treated as a doublet and provide the one-loop determinant on $\mathbb{S}^{2} \times \mathbb{S}^{1}$. Combining the contribution of the vector multiplet, the index is written\footnote{We can replace $|B_{J}|$ with $B_{J}$ in the formula \eqref{d1loop}. See the detail argument of \cite{Krattenthaler:2011da}.} as
\begin{align} 
&
\mathcal{I}_{\text{SQED}^{( \mathcal{CP} + )}} ( x, \alpha, w ) \notag \\ 
&=
\frac{1}{2} q^{- \frac{1}{8}}
\frac{( q; q^{2} )_{\infty}}{( q^{2}; q^{2} )_{\infty}}
\sum_{m \in \mathbb{Z}}
\left( q^{\frac{1}{2}} a ^{- \frac{1}{2}} w \right)^{m}
\left[
\frac{( a^{- \frac{1}{2}} q^{m + 1}; q )_{\infty}}{( a^{\frac{1}{2}} q^{m}; q )_{\infty}}
+
\frac{( - a^{- \frac{1}{2}} q^{m + 1}; q )_{\infty}}{( - a^{\frac{1}{2}} q^{m}; q )_{\infty}}
\right], 
\label{scib3sqed1}
\end{align}
where we define $m = B_{J}/2$ and $a = \alpha^{2} x^{2 \Delta}$. Note that we use $a$ with a different definition from the previous subsection. Also, $w := e^{i \theta_{J}}$ is a fugacity associated to gauged U$(1)_{J}$.

We express the dual XYZ as X$^{( + )}$Y$^{( + )}$Z because $X$ and $Y$ are coupled to the even-holonomy sector of the chosen U$(1)_{V}$ background gauge field.
Each field behaves as a single field on $\mathbb{RP}^{2} \times \mathbb{S}^{1}$ as explained in Table \ref{b3-type}. We can immediately obtain the index of X$^{( + )}$Y$^{( + )}$Z from its quiver diagram \eqref{quiverCP+} as
\begin{align} 
\mathcal{I}_{\text{X}^{( + )}\text{Y}^{( + )}\text{Z}} ( x, \tilde{\alpha}, \tilde{w} )
&=
q^{- \frac{1}{8}}
\frac
{( \tilde{a}^{\frac{1}{2}} \tilde{w}^{- 1} q^{\frac{1}{2}}, \tilde{a}^{\frac{1}{2}} \tilde{w} q^{\frac{1}{2}}, \tilde{a}^{- 1} q; q^{2} )_{\infty}}
{( \tilde{a}^{- \frac{1}{2}} \tilde{w} q^{\frac{1}{2}}, \tilde{a}^{- \frac{1}{2}} \tilde{w}^{- 1} q^{\frac{1}{2}}, \tilde{a}; q^{2} )_{\infty}},
\label{scib3xyz1}
\end{align}
where $\tilde{a} := \tilde{\alpha}^{- 2} x^{2 \Delta}$ and $\tilde{w} := e^{i \theta_{V}}$ is a fugacity associated to U$(1)_{V}$.

$\mathcal{N} = 2$ mirror symmetry between SQED$^{( \mathcal{CP} + )}$ and X$^{( + )}$Y$^{( + )}$Z is established 
\begin{align}
\mathcal{I}_{\text{SQED}^{( \mathcal{CP} + )}} ( x, \alpha, w )
=
\mathcal{I}_{\text{X$^{(+)}$Y$^{(+)}$Z}} ( x, \tilde{\alpha}, \tilde{w} )
\label{mirror3}
\end{align}
under identifications $\alpha = \tilde{\alpha}^{- 1}$ and $w = \tilde{w}$ which are concluded from the mirror map (Table \ref{mmap}). This is a quite new relation realizing mirror symmetry on the unorientable manifold.

\clearpage
\begin{align}
\xymatrix {
&
*++[o][F.]{\text{\scriptsize$B_{A}^{+}, \theta_A$}} 
&
\\
*+[F-]{\text{\scriptsize$
\begin{matrix}
Q \\
\tilde{Q}
\end{matrix}
$}}
\ar@{.>}[ur]<1.5mm>
\ar@{.>}[ur]<-1.5mm>
&
*++[o][F-]{\text{\scriptsize$\mc{CP}$}} 
\ar@{->}[l]<1.5mm>
\ar@{<-}[l]<-1.5mm>
\ar@{.}[d]^{\text{BF}}
&
\\
&
*++[o][F.]{\text{\scriptsize$B_{J}^{-}, \theta_J$}}
\\
&
\text{SQED}^{(\mc{CP}-)}
&
}
\hspace{-2em}
\begin{xy}
{(5,-15) \ar @{<=>} (15,-15)}
\end{xy}
\hspace{2em}
\xymatrix {
&
*++[o][F.]{\text{\scriptsize$\tilde{B}_{A}^{+}, \tilde{\theta}_A$}} 
\ar@{<.}[d]
\ar@{<.}[dl]
&
\\
*+[F-]{\text{$X$}} 
\ar@{.>}[dr]
&
*+[F-]{\text{$Y$}} 
\ar@{<.}[d]
&
*+[F-]{\text{$Z$}}
\ar@{<<.}[ul]
\\
&
*++[o][F.]{\text{\scriptsize$B_{V}^{-}, \theta_V$}}
& }
\notag \\[-1.75em]
&\hspace{-7.5em}
\text{X}^{(-)}
\text{Y}^{(-)}
\text{Z}
\label{quiverCP-} \\[-1em]
\notag
\end{align}

\paragraph{SQED$^{( \mathcal{CP} - )}$ vs X$^{( - )}$Y$^{( - )}$Z}
As the final example, we would like to mention SQED$^{( \mathcal{CP} - )}$ which is the SQED with the vector multiplet $V^{( \mathcal{CP} )}$ coupling to the fixed background gauge holonomy $B_{J}^{-}$ of U$(1)_{J}$ through the BF term. The quiver diagram of SQED$^{( \mathcal{CP} - )}$ is drawn on the lhs of \eqref{quiverCP-}. The matters $Q$ and $\tilde{Q}$ should be treated as a doublet as in SQED$^{( \mathcal{CP} + )}$, whereas a chosen $B_{J}^{-}$ produces the negative sign for the odd holonomy sector of the dynamical gauge field in summing up its configuration, which originated from the BF term \eqref{BFvalue}. This is an only difference from the previous case, therefore, we have the index of SQED$^{( \mathcal{CP} - )}$
\begin{align} 
&
\mathcal{I}_{\text{SQED}^{( \mathcal{CP} - )}} ( x, \alpha, w ) \notag \\ 
&=
\frac{1}{2} q^{- \frac{1}{8}}
\frac{( q; q^{2} )_{\infty}}{( q^{2}; q^{2} )_{\infty}}
\sum_{m \in \mathbb{Z}}
\left( q^{\frac{1}{2}} a^{- \frac{1}{2}} w \right)^{m}
\left[
\frac{( a^{- \frac{1}{2}} q^{m + 1}; q )_{\infty}}{( a^{\frac{1}{2}} q^{m}; q )_{\infty}}
-
\frac{( - a^{- \frac{1}{2}} q^{m + 1}; q )_{\infty}}{( - a^{\frac{1}{2}} q^{m}; q )_{\infty}} \right], 
\label{scib4sqed1}
\end{align}
where we repeat to use the variables defined in $\mathcal{I}_{\text{SQED}^{( \mathcal{CP} + )}}$.

We denote the theory dual to SQED$^{( \mathcal{CP} - )}$ as X$^{( - )}$Y$^{( - )}$Z characterized by the parity conditions shown in Table \ref{b3-type} with selecting the background gauge field $B_{V}^{-}$. This coupling leads to the situation where $X$ and $Y$ belong to the odd-holonomy sector of the matter multiplet on $\mathbb{RP}^{2} \times \mathbb{S}^{1}$, while $Z$ does not receive any effect. We quote the contributions of such chiral fields from the subsection \ref{Recipe} so that the index of X$^{( - )}$Y$^{( - )}$Z is obtained as
\begin{align} 
\mathcal{I}_{\text{X}^{( - )}\text{Y}^{( - )}\text{Z}} ( x, \tilde{\alpha}, \tilde{w} )
&=
q^{- \frac{1}{8}} \tilde{a}^{\frac{1}{2}}
\frac
{( \tilde{a}^{\frac{1}{2}} \tilde{w}^{- 1} q^{\frac{3}{2}}, \tilde{a}^{\frac{1}{2}} \tilde{w} q^{\frac{3}{2}}, \tilde{a}^{- 1} q; q^{2} )_{\infty}}
{( \tilde{a}^{- \frac{1}{2}} \tilde{w} q^{\frac{3}{2}}, \tilde{a}^{- \frac{1}{2}} \tilde{w}^{- 1} q^{\frac{3}{2}}, \tilde{a}; q^{2} )_{\infty}},
\label{scib4xyz1}
\end{align}
where $\tilde{a}$ is the same one defined in the previous example.

We declare $\mathcal{N} = 2$ mirror symmetry which connects SQED$^{( \mathcal{CP} - )}$ with X$^{( - )}$Y$^{( - )}$Z on $\mathbb{RP}^{2} \times \mathbb{S}^{1}$ as the nontrivial identity
\begin{align}
\mathcal{I}_{\text{SQED}^{( \mathcal{CP} - )}} ( x, \alpha, w )
=
\mathcal{I}_{\text{X$^{(-)}$Y$^{(-)}$Z}} ( x, \tilde{\alpha}, \tilde{w} )
\label{mirror4}
\end{align}
under identifications $\alpha = \tilde{\alpha}^{- 1}$ and $w = \tilde{w}$. We provide rigorous proof of the statements \eqref{mirror3} and \eqref{mirror4} by essencially employing the Ramanujan's summation formula \eqref{rama} and the specific product-to-sum identity of the elliptic theta functions \eqref{pts4443} step by step.


\bibliographystyle{common_supp_files/sonota/utphys}
\bibliography{common_supp_files/Ref}


\section{Conclusion and outlook}
We have derived two types of the consistent parity conditions imposed on all fields on $\mathbb{RP}^{2} \times \mathbb{S}^{1}$; one is just parity $\mathcal{P}$ acting as a set of \eqref{vecp1} and \eqref{ematp1}; the other is parity $\mathcal{P}$ accompanying charge conjugation $\mathcal{C}$ described in \eqref{vecp2} and \eqref{ematp2}. Then, the superconformal indices on $\mathbb{RP}^{2} \times \mathbb{S}^{1}$ have been computed under both pairs of the parity conditions. As an application of these indices, we check $\mathcal{N} = 2$ mirror symmetry with a single flavor on the unorientable manifold. As demonstrated in \cite{Tanaka:2014oda} where the parity conditions \eqref{vecp1} and \eqref{ematp1} without the BF term are considered, equivalence $\mathcal{I}_{\text{SQED}^{( \mathcal{P} )}} = \mathcal{I}_{^{\text{X}}_{\text{Y}}\text{Z}}$ for mirror symmetry can be proven essencially by the $q$-binomial theorem. On the other hand, we can verify mirror symmetry $\mathcal{I}_{\text{SQED}^{( \mathcal{CP} )}} = \mathcal{I}_{\text{XYZ}}$ under the parity conditions \eqref{vecp2} and \eqref{ematp2} with the Ramanujan's sum and the product-to-sum identity of the elliptic theta functions (see Appendix \ref{24mirrorsym} for details). Although its proof is rather simple, the resultant identity is quite nontrivial and new in the mathematical point of view. In the rest of this section, we mention some open problems and possibilities to generalize our argument.

We should comment on our variety of Abelian mirror symmetry on $\mathbb{RP}^{2} \times \mathbb{S}^{1}$ which can result from the revised index formula \eqref{d1loop} in the latest version of the paper. In the subsection \ref{Ptype} and \ref{CPtype}, we only consider Abelian mirror symmetry where the background gauge holonomy of U$(1)_A$ is restricted to even and realized it in four distinct ways depending on the parity conditions. In addition to them, we are suitably able to show other four types of Abelian mirror symmetry with U$(1)_A$ odd-holonomy $B_{A}^{-}$ and $\tilde{B}_{A}^{-}$ by applying the modified index formula \eqref{d1loop}, which we would omit here. One can easily check them in the level of the index following the same procedure presented above.

\paragraph{Open questions and outlook}
The naive issue involved in the map of the parity conditions is that we do not precisely understand the origin of the parity conditions for the XYZ model. To reveal this origin might be a crucial role to deeply understand the supersymmetric theories on unorientable manifolds.

There are many interesting extensions as future works. The simple one is to establish $\mathcal{N} = 2$ mirror symmetry with general $N_{f}$ flavors by  making use of $\mathcal{N} = 4$ mirror symmetry \cite{Kapustin:2011jm}. Furthermore, we did not perform the localization with non-Abelian gauge groups since we do not know the explicit form of a flat connection on $\mathbb{RP}^{2}$, namely, the Jacobian come from projecting the integration measure onto the Cartan subalgebra. We should try to overcome this point and derive more general formulas of the indices on $\mathbb{RP}^{2} \times \mathbb{S}^{1}$. The third one is to find the so-called {\it factorization} property of exact partition functions and indices on 3d compact curved manifolds \cite{Pasquetti:2011fj, Beem:2012mb, Hwang:2012jh, Taki:2013opa}. The fact that this is related to the solid-torus decomposition of a manifold leads to building blocks called {\it holomorphic blocks} whose specific combinations produce the partition functions and the indices. Physically, the holomorphic blocks are constructed by the inner products of some kinds of states which are defined on subspaces in decomposing the 3d manifolds. We expect that the blocks on $\mathbb{RP}^{2} \times \mathbb{S}^{1}$ can be described by the combinations of cross-cap states and boundary states on 2d submanifolds building the unorientable space. We would like to find the holomorphic blocks for our indices corresponding to the norms with such states, or, an appropriate way of gluing the surface of a solid torus. Moreover, the brane construction in string theory \cite{deBoer:1997ka} is an important perspective to discuss mirror symmetry on $\mathbb{RP}^{2} \times \mathbb{S}^{1}$. It may be possible to resolve the questions mentioned above from the brane construction viewpoints. Finally, we aim to generalize the identities \eqref{mirror3} and \eqref{mirror4} in mathematical sense as the discussions about the $q$-binomial theorem \cite{Tanaka:2014oda}. We guess that there is a more general formula which reproduces them by taking a certain combination of parameters. This is a very fascinating aspect of our results that physics provides new insights in mathematics, and we hope that the generalizations of our mathematical arguments may be fed back to physical veiwpoints.

\bibliographystyle{common_supp_files/sonota/utphys}
\bibliography{common_supp_files/Ref}

\acknowledgments{We would like to thank Heng-Yu Chen, Koji Hashimoto, Kazuo Hosomichi, Norihiro Iizuka, Yosuke Imamura, Taro Kimura, Yousuke Ohyama, Yuji Sugawara, Masato Taki, Seiji Terashima, and Satoshi Yamaguchi. This research was supported by the RIKEN iTHES Project. The work of H.M. and T.M. was supported in part by the JSPS Research Fellowship for Young Scientists.}
\appendix

\section{Localization calculus for new sectors}\label{LOCALIZATION}
\paragraph{Locus}
Surprisingly, we can construct a monopole even on $\Rp^2$.
However, there is one constraint.
Locus configuration for the vector multiplet in \eqref{vecp2} is characterized by
\begin{align}
A = B A_{\text{mon}} + \frac{\theta_\pm}{2 \pi} d \compd,
\quad
\si = - \frac{B}{2}
\qquad
(\theta_\pm \sim \theta_\pm + 2 \pi),
\label{locus}
\end{align}
where $A_{\text{mon}}$ is the Dirac monopole configuration,
\begin{align}
A_{\text{mon}}
=
\left\{ \begin{array}{ll}
\frac{1}{2} (+1 - \cos \compe) d \compt, &\quad \compe \in [0 , \pi), \\[.5em]
\frac{1}{2} (-1 - \cos \compe) d \compt, &\quad \compe \in ( 0 , \pi ].\\
\end{array} \right.
\end{align}
The upper gauge field is connected to the lower one by the gauge transformation
\begin{align}
g (\compt) = e^{i B \compt}.
\end{align}
In order to make the gauge transformation single-valued on $\Rp^2$, we should set the constraint $g(\pi) = g(0)$, and the monopole charge $B$ is quantized to be even integers
\begin{align}
B \in 2 \mathbb{Z}.
\end{align}
This condition is different from the usual Dirac quantization condition $B \in \mathbb{Z}$.
In addition to it, we should restrict $\theta$ as
\begin{align}
\theta_+ = 0,
\quad
\theta_- = \pi,
\label{tlocus}
\end{align}
in order to make it consistent with the $A_\compd$ condition in \eqref{vecp2} mod $2\pi$.
For matter multiplet, 
there is no non-trivial locus with our matter Lagrangian.

\paragraph{One-loop determinant for vector multiplet}
First, we focus on the contribution from $V^{(\mc{CP})}$.
Most of the calculation can be achieved parallel to the argument in \cite{Tanaka:2014oda}.
The one-loop determinant is given by the product of
\begin{align}
\frac{\text{(Unpaired fermionic eigenvalue)}}{\text{(Unpaired bosonic eigenvalue)}}
=
\frac{2 \beta_2 ( - j_3^f +1) +2 \pi i  n }
{2\beta_2 (+ j_3^b +1) +2 \pi i  n},
\end{align}
where $\beta_2$ is related to $x$ in \eqref{sci} as $e^{-\beta_2} = x$.
$j_3$ is restricted as $j_3^b \geq 1$ and $j_3^f \leq 0$ in order to make the corresponding eigenmodes regular around $\compe = 0$,
and
\begin{align}
e^{ij_3 \pi} = +1,
\label{count}
\end{align}
where it is a consequence of the conditions in \eqref{vecp2}.
This constraint \eqref{count} is a counterpart to (A.27) in \cite{Tanaka:2014oda}.
It gives
\begin{align}
j_3^b = 2, 4, \dots,
\quad
j_3^f =  0, -2, -4, \dots
\end{align}
or equivalently, $j_3^b = 2k+2$ and $j_3^f = -2k$ for $k\geq 0$. Thus, we get
\begin{align}
\mc{Z}_{\text{1-loop}}^{(\mc{CP})}
&=
\prod_{n \in \mathbb{Z}}
\prod_{k \geq 0}
\frac{2\beta_2 ( 2k +1) +2 \pi i  n }
{2\beta_2 (2k +3) +2 \pi i  n}
\notag \\
&=
x^{-\frac{1}{4}} \frac{(x^2 ; x^4)_\infty}{(x^4 ; x^4)_\infty}.
\end{align}
To get the final expression, we use the same regularization scheme used in \cite{Tanaka:2014oda}.

\paragraph{One-loop determinant for matter multiplet}
Subsequently, we consider the one-loop determinant for $\Phi_d$, or more precisely, we show the formula
\begin{align}
\xymatrix{
*++[o][F.]{\text{\scriptsize$A^\pm_1 , \theta_1$}} 
\ar@{<.}[r]<1mm>
\ar@{<.}[r]<-1mm>_{\underbrace{}_{\ch_1}}
&
*+[F-]{^{\Phi_1}_{\Phi_2}}
\ar@{.>}[r]<1mm>
\ar@{<.}[r]<-1mm>_{\underbrace{}_{\ch_2}}
&
*++[o][F.]{\text{\scriptsize$B_2 , \theta{_2}{_\pm}$}} 
}
&=
e^{i \frac{\ch_1 \ch_2 A_{1}^{\pm} \theta{_2}{_\pm}}{2}}
 \Big(  x^{1-\Delta} e^{- i \ch_1 \theta_1} 
\Big)^{\frac{|\ch_2 B_2|}{2}}
\frac{\Big(
e^{-i ( \ch_1 \theta{_1} +\ch_2 \theta{_2}{_\pm}) }  
x^{|\ch_2 B_2|+(2-\Delta)} ; x^2 \Big)_\infty}
{\Big(
e^{+i (\ch_1 \theta{_1}+\ch_2 \theta{_2}{_\pm}  )}  
x^{|\ch_2 B_2|+(0+\Delta)} ; x^2\Big)_\infty}.
\end{align}
First of all, we redefine the matter multiplets as
\begin{align}
\tilde{\Phi}_{1,2}
:= e^{i \int^x \ch_1 A_1^\pm} \Phi_{1,2},
\end{align}
then, we can remove the background gauge field $A_1^\pm$ from the Lagrangian.
As a next step, we use the doubling trick and define a matter multiplet $\tilde{\Phi}$ on virtual $\Sp^2 \times \Sp^1$ as
\begin{align}
\tilde{\Phi} (\compe, \compt, \compd)
=
\left\{ \begin{array}{ll}
\tilde{\Phi}_1 (\compe, \compt, \compd),& \compe \in [0 , \frac{\pi}{2}],  \\[.5em]
 e^{i \oint_\g \ch_1 A_1^\pm}  \tilde{\Phi}_2 (\compe, \compt, \compd), & \compe \in [\frac{\pi}{2},\pi],  \\
\end{array} \right.
\label{dtos}
\end{align}
where $\g$ is the non-trivial cycle of $\Rp^2$, and get the Lagrangian with covariant derivative
\begin{align}
\mathcal{D} = dx^\mu \nabla_\mu 
- i \ch_2 B_2 A_{\text{mon}}
- i \frac{\ch_1  \theta_1 + \ch_2 \theta{_2}{_\pm}}{2 \pi} d\compd.
\label{cov}
\end{align}
Therefore, we can recycle the one-loop determinant for a single matter multiplet on $\Sp^2 \times \Sp^1$ with covariant derivative $\mathcal{D} = dx^\mu \nabla_\mu - i \ch B A_{\text{mon}} - i \frac{\ch \theta}{2 \pi} d \compd$  \cite{Imamura:2011su}
\begin{align}
\mc{Z}_{\text{1-loop}}^{\Sp^2 \times \Sp^1} (\ch \theta, \ch B)
=
\Big(  x^{1- \Delta}e^{- i \ch \theta }  \Big)^{\frac{|\ch B|}{2}}
\frac{(e^{- i \ch \theta} x^{|\ch B|+(2-\Delta)} ; x^2)_\infty}{(e^{+i \ch \theta } x^{|\ch B|+ ( 0+\Delta)} ; x^2)_\infty}
.
\end{align}
Naively, the expected one-loop determinant for the matter in \eqref{dtos} with \eqref{cov} is
\begin{align}
e^{i \frac{\ch_1 \ch_2 A_{1}^{\pm} \theta{_2}{_\pm}}{2}}
\Big(  x^{1- \Delta}e^{- i (\ch_1 \theta_1 + \ch_2 \theta{_2}{_\pm}) }  \Big)^{\frac{|\ch_2 B_2|}{2}}
\frac{(e^{- i (\ch_1 \theta_1 + \ch_2 \theta{_2}{_\pm})} x^{|\ch_2 B|+(2-\Delta)} ; x^2)_\infty}{(e^{+i (\ch_1 \theta_1 + \ch_2 \theta{_2}{_\pm})} x^{|\ch_2 B_2|+ ( 0+\Delta)} ; x^2)_\infty}
.
\end{align}
The first exponential factor might arise from the BF coupling between the background gauge fields for two U$(1)$ global symmetries.
It is almost correct but not perfect one.
We should be careful about the regularization of the prefactor called the Casimir energy.
In \cite{Imamura:2011su}, for example, the Casimir energy is derived after a certain derivative operation with respect to $z=e^{i\theta}$.
In our situation, however, we cannot do that because $e^{i \theta{_2}{_\pm}} = (-1)^\pm$ and it is not a continuous variable. 
We propose that the precise regularization is just dropping the $e^{-i \ch_2 \theta{_2}{_\pm}}$ factor from the Casimir part.
In fact, the only problematic case is for $\ch_2 \theta{_2} = \pi$ mod $2 \pi$, and, in this case, the resultant one-loop determinant can be expressed by
\begin{align}
\prod \frac{2 \sinh ( i \frac{\pi}{2} + \dots)}{2 \sinh ( i \frac{\pi}{2} + \dots)}
=
\prod \frac{2 \cosh ( \dots)}{2 \cosh (\dots)}.
\end{align}
Therefore, our result seems to be correct.
In addition, we confirm validity of our regularization through the agreements of expected identities expected from 3d mirror symmetry in the main context.

\paragraph{Classical BF contribution}
We derive here the formula \eqref{BFvalue}.
What we consider here is the value for \eqref{BFt} around the loci in \eqref{PBG} and \eqref{CPBG}, and it becomes
\begin{align}
\frac{i}{2 \pi} S_{\text{BF}} (A^{(\mc{P})},A^{(\mc{CP})})
=
\frac{i}{2 \pi} \int_{\Rp^2 \times \Sp^1}
A^{(\mc{P})} \wedge F^{(\mc{CP})}.
\label{ApBF}
\end{align}
We should be careful to calculate it as noted in \cite{Kim:2009wb, Gang:2009wy, Imamura:2011su}.
One way is as follows.
First, we consider the four-manifold $\Rp^2 \times \mathbb{D}^2$ with boundary $\Rp^2 \times \Sp^1$ and extend the 3d gauge field $A$ to the 4d gauge field $\tilde{A}$.
Then, the action \eqref{ApBF} can be rewritten by the following 4d integral:
\begin{align}
\frac{i}{2\pi}
\int_{\Rp^2 \times \mathbb{D}^2}
\tilde{F}^{(\mc{P})} 
\wedge
\tilde{F}^{(\mc{CP})}
&=
\frac{i}{2\pi}
\Big(
\int_{\Rp^2}
\tilde{F}^{(\mc{P})} 
\int_{\mathbb{D}^2}
\tilde{F}^{(\mc{CP})}
+
\int_{\Rp^2}
\tilde{F}^{(\mc{CP})} 
\int_{\mathbb{D}^2}
\tilde{F}^{(\mc{P})}
\Big)
\notag \\
&=
\frac{i}{2\pi}
\Big(
\int_{\Rp^2}
d (B^\pm A_{\text{flat}})
\cdot
\theta_\pm
+
\int_{\Rp^2}
d (B A_{\text{mon}})
\cdot
\theta
\Big).
\end{align}
In order to compute the integral over $\Rp^2$, 
we use a trick as follows:
\begin{align}
&\int_{\Rp^2} d A
=
\int_{\pa \Rp^2} A
\sim
\oint_{\g + \g}A
=
2 \oint_\g A,
\end{align}
where $\g$ is a non-trivial cycle. Then, we get
\begin{align}
&\int_{\Rp^2} 
d (B^\pm  A_{\text{flat}})
=
2 B^\pm \oint_\g  A_{\text{flat}},
\\
&\int_{\Rp^2}
d (B A_{\text{mon}})
=
2 B \oint_\g  A_{\text{mon}}^{\text{upper}} \Big|_{\compe = \pi/2}
=
2 B \int_0^\pi \frac{1}{2}d \compt
=
B \pi,
\end{align}
therefore, we arrive at
\begin{align}
e^{\frac{i}{2 \pi} S_{\text{BF}} (A^{(\mc{P})},A^{(\mc{CP})})}
&=
e^{
\frac{i}{2\pi}
\Big(
2 B^\pm \oint_\g  A_{\text{flat}}
\cdot
\theta_\pm
+
B \pi
\cdot
\theta
\Big)
}
\notag \\
&=
(
-1
)^{{B^\pm \theta_\pm}/{\pi} }
\times
\Big( e^{i \theta}
\Big)
^{B/2},
\end{align}
where we used the feature of $A_{\text{flat}}$
\begin{align}
e^{i \oint_\g A_{\text{flat}}}
=
(-1).
\end{align}

\bibliographystyle{common_supp_files/sonota/utphys}
\bibliography{common_supp_files/Ref}


\section{Proof of mirror symmetry}\label{24mirrorsym}

\subsection*{Mathematical preliminaries}\label{maths}
The superconformal index is written as the polynomial of fugacities associated to symmetries which commute with the selected supercharges. Alternatively, it can be summarized in the form of infinite products written as special functions. We would like to gather mathematical ingredients used to express our indices compactly and indicate mirror symmetry as equivalence between the indices.

First of all, we make a list of definitions and formulas of the hypergeometric series \cite{Gasper} which we use in rewriting the indices as simple forms.
\begin{itemize}
\item $q$-shifted factorial ($q$-Pochhammer symbol):
\begin{align} 
( z ; q )_{n} = \left\{
	\begin{aligned}
	& 1 && \mbox{for } n = 0, \\[.75em] 
	& \prod_{k = 0}^{n - 1} ( 1- z q^{k} ) && \mbox{for } n \geq 1, \\ 
	& \prod_{k = 1}^{- n} ( 1- z q^{- k} )^{- 1} && \mbox{for } n \leq - 1, 
	\end{aligned} \right.
\label{qshifted}
\end{align}
where $z$ and $q$ are complex numbers, and $( z ; q )_{\infty} := \lim_{n \to \infty} ( z ; q )_{n}$ with $0 < | q | < 1$. For simplicity, we use the shorthand notation
\begin{align}
( z_{1}, z_{2}, \cdots, z_{r} ; q )_{n} := ( z_{1} ; q )_{n} ( z_{2} ; q )_{n} \cdots ( z_{r}; q )_{n}.
\end{align}
In addition, we note useful formulas to change the power of $q$ in the $q$-shifted factorial,
\begin{align}
	\begin{aligned}
	( z, z q; q^{2} )_{\infty} &= ( z; q )_{\infty}, \\ 
	( z; q^{2} )_{\infty} &= ( z^{\frac{1}{2}}, - z^{\frac{1}{2}}; q )_{\infty}. 
	\end{aligned}
\label{qdecomp}
\end{align}

\item Basic hypergeometric series:
\begin{align} 
{}_{r}\varphi_{s} \left( \alpha_{1}, \alpha_{2}, \cdots, \alpha_{r} ; \beta_{1}, \cdots, \beta_{s} ; q, z \right)
= \sum_{n = 0}^{\infty}
\frac{( \alpha_{1}, \alpha_{2}, \cdots, \alpha_{r}; q )_{n}}{( \beta_{1}, \cdots, \beta_{s}; q )_{n}}
\frac{z^{n}}{( q; q)_{n}} \left\{ ( - 1 )^{n} q^{\frac{1}{2} n ( n - 1 )} \right\}^{1 + s - r}.
\label{bhs}
\end{align}
The convergence radius is $\infty$, $1$, or $0$ for $r - s < 1$, $r - s = 1$, or $r - s > 1$, respectively.

\item General bilateral basic hypergeometric series:
\begin{align} 
{}_{r}\psi_{s} ( \alpha_{1}, \cdots, \alpha_{r}; \beta_{1}, \cdots, \beta_{s}; q, z )
= \sum_{n = - \infty}^{\infty}
\frac{( \alpha_{1}, \cdots, \alpha_{r}; q )_{n}}{( \beta_{1}, \cdots, \beta_{s}; q )_{n}}
z^{n}
\left\{ ( - 1 )^{n} q^{\frac{1}{2} n ( n - 1 )} \right\}^{s - r}.
\label{gbbhs}
\end{align}
Notice that, assuming $|q| < 1$, this series converges on
\begin{align*}
\left\{ \begin{aligned}
	&R < |z| &&\text{ for } s > r, \\ 
	&R < |z| < 1 &&\text{ for } s = r, 
	\end{aligned} \right.
\end{align*}
where $R := | \beta_{1} \cdots \beta_{s} / \alpha_{1} \cdots \alpha_{r} |$, while it becomes a divergent series around $z = 0$ for $s < r$.

\item $q$-binomial theorem:
\begin{equation} 
{}_1\varphi_0(a;-;q,z)=\frac{(az;q)_\infty}{(z;q)_\infty}
\label{qbinomial}
\end{equation}
for $|z|<1$. In the $q \to 1$ limit, this reduces to the usual binomial theorem.

\item Ramanujan's summation formula:
\begin{align} 
{}_{1}\psi_{1} ( a; b; q, z ) = \frac{( q, b/a, a z, q/ az; q )_{\infty}}{( b, q/a, z, b/az; q )_{\infty}}
\label{rama}
\end{align}
with $|b/a| < |z| < 1$. Notice that when we take an appropriate limit for each parameter, this formula reproduces the $q$-binomial theorem \eqref{qbinomial}.
\end{itemize}
The readers interested in detailed mathematical aspects of these series and associated formulas are offered to see the literature \cite{Gasper}.

To verify mirror symmetry in terms of the indices later, we will utilize the elliptic theta functions defined as follows \textmd{\cite{Whittaker}}:
\begin{align} 
\vartheta_{4} ( z )
=
\vartheta_{4} ( z, \tau )
&=
\sum_{n = - \infty}^{\infty} ( - 1 )^{n} q^{\frac{n^{2}}{2}} x^{n},
\label{et4} \\ 
\vartheta_{3} ( z )
=
\vartheta_{3} ( z, \tau )
&=
\sum_{n = - \infty}^{\infty} q^{\frac{n^{2}}{2}} x^{n},
\label{et3} \\ 
\vartheta_{2} ( z )
=
\vartheta_{2} ( z, \tau ) 
&=
\sum_{n = - \infty}^{\infty} q^{\frac{1}{2} \left( n - \frac{1}{2} \right)^{2}} x^{n - \frac{1}{2}},
\label{et3} \\ 
\vartheta_{1} ( z )
=
\vartheta_{1} ( z, \tau )
&=
i \sum_{n = - \infty}^{\infty} ( - 1 )^{n} q^{\frac{1}{2} \left( n - \frac{1}{2} \right)^{2}} x^{n - \frac{1}{2}},
\label{et1} 
\end{align}
where $q = e^{2 \pi i \tau}$ and $x = e^{2 \pi i z}$. We do not say explicitly the dependence on $\tau$ in these elliptic theta functions otherwise we state. They are, of course, elliptic functions and satisfy useful formulas listed below which we can check from their definitions.
\begin{itemize} 
\item Inversion formulas:
\begin{align}
\vartheta_{4} ( z ) &= \vartheta_{4} ( - z ),
\label{inversion4} \\ 
\vartheta_{3} ( z ) &= \vartheta_{3} ( - z ).
\label{inversion3} 
\end{align}

\item Periodicity properties:
\begin{align}
\vartheta_{4} ( z + 1 ) &= \vartheta_{4} ( z ),
&&
\vartheta_{4} ( z + \tau ) = - q^{- \frac{1}{2}} x^{- 1} \vartheta_{4} ( z ),
\label{period4} \\ 
\vartheta_{3} ( z + 1 ) &= \vartheta_{3} ( z ),
&&
\vartheta_{3} ( z + \tau ) = q^{- \frac{1}{2}} x^{- 1} \vartheta_{3} ( z ).
\label{period3} 
\end{align}

\item Jacobi's triple product identities:
\begin{align} 
\vartheta_{4} ( z )
&= ( q, q^{\frac{1}{2}} x, q^{\frac{1}{2}}/x; q )_{\infty},
\label{triple4} \\ 
\vartheta_{3} ( z )
&= ( q, - q^{\frac{1}{2}} x, - q^{\frac{1}{2}}/x; q )_{\infty}.
\label{triple3}
\end{align}
\end{itemize}
The other theta functions $\vartheta_{1}$ and $\vartheta_{2}$ fulfil similar properties which we do not show here.

Next, we show the product-to-sum identity related to $\vartheta_{3}$ and $\vartheta_{4}$. The combinations of the elliptic theta functions with different arguments exhibit hundreds of product-to-sum identities as trigonometric functions. Because one of them has an important role to prove our mirror symmetry, we would like to derive it here in the generic form. To do it, we note that the following identity holds:
\begin{align}
\vartheta_{4} ( z - w, 2 \tau ) \vartheta_{4} ( z + w, 2 \tau ) = \vartheta_{4} ( z, \tau ) \vartheta_{3} ( w, \tau ).
\label{vt4443}
\end{align}
\begin{proof}
From the definition of $\vartheta_{4}$,
\begin{align*}
& \vartheta_{4} ( z - w, 2 \tau ) \vartheta_{4} ( z + w, 2 \tau ) \\ 
&= \left( \sum_{n \in \mathbb{Z}} \exp \left[ \pi i n + 2 \pi i n^2 \tau + 2 \pi i n ( z - w ) \right] \right)
\left( \sum_{m \in \mathbb{Z}} \exp \left[ \pi i m + 2 \pi i m^2 \tau + 2 \pi i m ( z + w ) \right] \right) \\ 
&= \sum_{n \in \mathbb{Z}} \sum_{m \in \mathbb{Z}}
\exp \left[ \pi i ( n + m ) + 2 \pi i ( n^2 + m^2 ) \tau + 2 \pi i n ( z - w ) + 2 \pi i m ( z + w ) \right] \\ 
&= \sum_{n \in \mathbb{Z}} \sum_{m \in \mathbb{Z}}
\exp \left[ \pi i ( n + m )^2 \tau + \pi i ( n - m )^2 \tau +
\pi i ( n + m ) z - \pi i ( n - m ) w + \pi i ( n + m ) \right] \\ 
&= \left( \sum_{N \in \mathbb{Z}} \exp \left[ \pi i N^2 \tau + \pi i N z + \pi i N \right] \right)
\left( \sum_{ - M \in \mathbb{Z}} \exp \left[ \pi i M^2 \tau +\pi i M w \right] \right) \\ 
&= \vartheta_{4} ( z, \tau ) \vartheta_{3} ( w, \tau ), 
\end{align*}
where $N := n + m$ and $M := n - m$.
\end{proof}
We think of this as the formula to recombine the elliptic theta functions accompanying the scale of $\tau$ in the way that an argument which appears with the same sign on the left-hand side (i.e., $z$ in \eqref{vt4443}) becomes one of $\vartheta_{4}$ on the right-hand side, and an argument which appears with the opposite sign on the left-hand side (i.e., $w$ in \eqref{vt4443}) becomes one of $\vartheta_{3}$ on the right-hand side.

Then, we rewrite the left-hand side of \eqref{vt4443} as the following product-to-sum identity:
\begin{eqnarray} 
\framebox[12.5cm][c]{\parbox{11.5cm}{
	\begin{eqnarray*}
	\vartheta_{4} ( z - w, 2 \tau ) \vartheta_{4} ( z + w, 2 \tau )
	=
	\frac{1}{2}
	\left\{
	\vartheta_{4} ( z, \tau ) \vartheta_{3} ( w, \tau )
	+
	\vartheta_{3} ( z, \tau ) \vartheta_{4} ( w, \tau )
	\right\}.
	\end{eqnarray*} }}
	\label{pts4443}
\end{eqnarray}
This formula turns to be a core to realize mirror symmetry under the new parity conditions.
\begin{proof}
Dividing $\vartheta_{4} ( z \mp w )$ into its sum by means of the inversion formulas \eqref{inversion4},
\begin{align*}
\vartheta_{4} ( z - w, 2 \tau )
&=
\frac{1}{2}
\left\{ \vartheta_{4} ( z - w, 2 \tau ) + \vartheta_{4} ( - z + w, 2 \tau ) \right\}, \\ 
\vartheta_{4} ( z + w, 2 \tau )
&=
\frac{1}{2}
\left\{ \vartheta_{4} ( z + w, 2 \tau ) + \vartheta_{4} ( - z - w, 2 \tau ) \right\}. 
\end{align*}
We substitute them into the lhs of \eqref{vt4443}, then applying \eqref{vt4443} again to each term,
\begin{align*}
\vartheta_{4} ( z - w, 2 \tau ) \vartheta_{4} ( z + w, 2 \tau )
&=
\frac{1}{4}
\left\{
\vartheta_{4} ( z - w, 2 \tau ) \vartheta_{4} ( z + w, 2 \tau )
+
\vartheta_{4} ( - z + w, 2 \tau ) \vartheta_{4} ( z + w, 2 \tau )
\right. \\ 
&\left. \hspace{2em} +
\vartheta_{4} ( z - w, 2 \tau ) \vartheta_{4} ( - z - w, 2 \tau )
+
\vartheta_{4} ( - z + w, 2 \tau ) \vartheta_{4} ( - z - w, 2 \tau )
\right\} \\ 
&=
\frac{1}{4}
\left\{
\vartheta_{4} ( z, \tau ) \vartheta_{3} ( w, \tau )
+
\vartheta_{3} ( z, \tau ) \vartheta_{4} ( w, \tau )
\right. \\ 
&\left. \hspace{2em} +
\vartheta_{3} ( z, \tau ) \vartheta_{4} ( - w, \tau )
+
\vartheta_{4} ( - z, \tau ) \vartheta_{3} ( w, \tau )
\right\} \\ 
&=
\frac{1}{2}
\left\{ \vartheta_{4} ( z, \tau ) \vartheta_{3} ( w, \tau ) + \vartheta_{3} ( z, \tau ) \vartheta_{4} ( w, \tau ) \right\}, 
\end{align*}
where we used the inversion formula \eqref{inversion4} in the last line.
\end{proof}


\subsection*{$\text{SQED}^{( \mathcal{P} + )}$ vs ${^{\text{X}}_{\text{Y}}}^{(+)}\text{Z}$}
Let us turn to proving $\mathcal{N} = 2$ mirror symmetry stated in Section \ref{Mirrorsym}. We start with the case of the fixed background Wilson line phase $\theta_{J +}$ in the subsection \ref{Ptype} where the index of the SQED$^{( \mathcal{P} + )}$ is given by
\begin{align} 
&
\mathcal{I}_{\text{SQED}^{( \mathcal{P} + )}} ( x, \alpha, B_{J} ) \notag \\ 
&=
q^{\frac{1}{8}}
\frac{( q^{2} ; q^{2} )_{\infty}}{( q ; q^{2} )_{\infty}}
\oint_{C_0} \frac{d z}{2 \pi i z} z^{s}
\left\{ a^{- \frac{1}{4}}
\frac{( z^{- 1} a^{\frac{1}{2}} q^{\frac{1}{2}}, z a^{\frac{1}{2}} q^{\frac{1}{2}}; q^{2} )_{\infty}}
{( z a^{- \frac{1}{2}} q^{\frac{1}{2}}, z^{- 1} a^{- \frac{1}{2}} q^{\frac{1}{2}} ; q^{2} )_{\infty}}
+
a^{\frac{1}{4}}
\frac{( z^{- 1} a^{\frac{1}{2}} q^{\frac{3}{2}}, z a^{\frac{1}{2}} q^{\frac{3}{2}} ; q^{2} )_{\infty}}
{( z a^{- \frac{1}{2}} q^{\frac{3}{2}}, z^{- 1} a^{- \frac{1}{2}} q^{\frac{3}{2}} ; q^{2} )_{\infty}} \right\}. 
\label{scib1sqed2}
\end{align}
As done in \cite{Krattenthaler:2011da, Kapustin:2011jm}, we pick up the set of poles in the integration depending on the value of $s$: the poles inside the unit circle for $s > 0$ and outside the unit circle for $s < 0$. Therefore, the poles in \eqref{scib1sqed2} which we should take into account can be summarized as
\begin{align} 
z = \left\{
	\begin{aligned}
	& \left( a^{- \frac{1}{2}} q^{\frac{1}{2} + 2 j} \right)^{\text{sign}(s)} \hspace{1em} \text{for the 1st term}, \\ 
	& \left( a^{- \frac{1}{2}} q^{\frac{3}{2} + 2 j} \right)^{\text{sign}(s)} \hspace{1em} \text{for the 2nd term} 
	\end{aligned}
\right.
\label{zpoles}
\end{align}
with $j = 0, 1, 2, \cdots$. The residues of \eqref{zpoles} in the index are evaluated as
\begin{align} 
&
\mathcal{I}_{\text{SQED}^{( \mathcal{P} + )}} \notag \\ 
&=
q^{\frac{1}{8}}
\frac{( q^{2} ; q^{2} )_{\infty}}{( q ; q^{2} )_{\infty}}
\left\{
a^{- \frac{1}{4}}
\sum_{j \geq 0}
\frac{( a q^{- 2 j}, q^{1 + 2 j} ; q^{2} )_{\infty}}{( a^{- 1} q^{1 + 2 j}, q^{2} ; q^{2} )_{\infty}}
\frac{( a^{- \frac{1}{2}} q^{\frac{1}{2} + 2 j} )^{| s |}}{( q^{- 2 j} ; q^{2} )_{j}} \right. \notag \\ 
&\hspace{7em} + \left.
a^{\frac{1}{4}}
\sum_{j \geq 0}
\frac{( a q^{- 2 j}, q^{3 + 2 j} ; q^{2} )_{\infty}}{( a^{- 1} q^{3 + 2 j}, q^{2} ; q^{2} )_{\infty}}
\frac{( a^{- \frac{1}{2}} q^{\frac{3}{2} + 2 j} )^{| s |}}{( q^{- 2 j} ; q^{2} )_{j}} \right\} \notag \\ 
&=
q^{\frac{1}{8}}
\frac{( q^{2} ; q^{2} )_{\infty}}{( q ; q^{2} )_{\infty}}
\left\{
a^{- \frac{1}{4}}
\left( a^{- \frac{1}{2}} q^{\frac{1}{2}} \right)^{| s |}
\frac{( a, q ; q^{2} )_{\infty}}{( a^{- 1} q, q^{2} ; q^{2} )_{\infty}}
\sum_{j \geq 0}
\frac{( a^{- 1} q^{2}, a^{- 1} q ; q^{2} )_{j}}{( q ; q^{2} )_{j}} \frac{a^{j} q^{2 |s| j}}{( q^{2} ; q^{2} )_{j}} \right. \notag \\ 
&\hspace{7em} + \left.
a^{\frac{1}{4}}
\left( a^{- \frac{1}{2}} q^{\frac{3}{2}} \right)^{| s |}
\frac{( a, q^{3} ; q^{2} )_{\infty}}{( a^{- 1} q^{3}, q^{2} ; q^{2} )_{\infty}}
\sum_{j \geq 0}
\frac{( a^{- 1} q^{2}, a^{- 1} q^{3} ; q^{2} )_{j}}{( q^{3} ; q^{2} )_{j}} \frac{a^{j} q^{2 |s| j}}{( q^{2} ; q^{2} )_{j}} \right\} \notag \\ 
&=
q^{\frac{1}{8}}
\frac{( q^{2} ; q^{2} )_{\infty}}{( q ; q^{2} )_{\infty}}
\left\{
a^{- \frac{1}{4}} \left( a^{- \frac{1}{2}} q^{\frac{1}{2}} \right)^{| s |}
\frac{( a, q ; q^{2} )_{\infty}}{( a^{- 1} q, q^{2} ; q^{2} )_{\infty}}
\ {}_{2}\varphi_{1} ( a^{- 1} q^{2}, a^{- 1} q ; q ; q^{2}, a q^{2 |s|} ) \right. \notag \\ 
&\hspace{7em} + \left.
a^{\frac{1}{4}}
\left( a^{- \frac{1}{2}} q^{\frac{3}{2}} \right)^{| s |}
\frac{( a, q^{3} ; q^{2} )_{\infty}}{( a^{- 1} q^{3}, q^{2} ; q^{2} )_{\infty}}
\ {}_{2}\varphi_{1} ( a^{- 1} q^{2}, a^{- 1} q^{3} ; q^{3} ; q^{2}, a q^{2 |s|} ) \right\}, 
\label{scib1sqed3}
\end{align}
where we use the basic hypergeometric series \eqref{bhs} with $r = 2$, $s = 1$.

On the dual side, the index of ${^{\text{X}}_{\text{Y}}}^{(+)}\text{Z}$ is
\begin{align} 
\mathcal{I}_{{^{\text{X}}_{\text{Y}}}^{(+)}\text{Z}} ( x, \tilde{\alpha}, B_{V} )
=
q^{\frac{1}{8}} \tilde{a}^{- \frac{1}{4}}
\left( \tilde{a}^{- \frac{1}{2}} q^{\frac{1}{2}} \right)^{| \tilde{s} |}
\frac{( \tilde{a}^{- \frac{1}{2}} q^{1 + |\tilde{s}|} ; q )_{\infty}}{( \tilde{a}^{\frac{1}{2}} q^{|\tilde{s}|} ; q )_{\infty}}
\frac{( \tilde{a} ; q^{2} )_{\infty}}{( \tilde{a}^{- 1} q ; q^{2} )_{\infty}}.
\label{scib1xyz2}
\end{align}
We aim to rearrange the power of $q$ of the first fraction in \eqref{scib1xyz2} to $q^{2}$ by utilizing the $q$-binomial theorem \eqref{qbinomial}. Specifically, we re-express it as
\begin{align} 
&
\frac{( \tilde{a}^{- \frac{1}{2}} q^{1 + |\tilde{s}|} ; q )_{\infty}}{( \tilde{a}^{\frac{1}{2}} q^{|\tilde{s}|} ; q )_{\infty}} \notag \\ 
&=
\ {}_{1}\varphi_{0} ( \tilde{a}^{-1} q; - ; q, \tilde{a}^{\frac{1}{2}} q^{|\tilde{s}|} ) \notag \\ 
&=
\sum_{j \geq 0}
\frac{( \tilde{a}^{-1} q; q )_{j}}{( q; q )_{j}} \left( \tilde{a}^{\frac{1}{2}} q^{|\tilde{s}|} \right)^{j} \notag \\ 
&=
\sum_{m \geq 0}
\frac{( \tilde{a}^{-1} q; q )_{2m}}{( q; q )_{2m}} \left( \tilde{a}^{\frac{1}{2}} q^{|\tilde{s}|} \right)^{2 m}
+
\sum_{m \geq 0}
\frac{( \tilde{a}^{-1} q; q )_{2m + 1}}{( q; q )_{2m + 1}} \left( \tilde{a}^{\frac{1}{2}} q^{|\tilde{s}|} \right)^{2 m + 1} \notag \\ 
&=
\sum_{m \geq 0}
\frac{( \tilde{a}^{-1} q, \tilde{a}^{-1} q^{2}; q^{2} )_{m}}{( q, q^{2}; q^{2} )_{m}} \left( \tilde{a} q^{2 |\tilde{s}|} \right)^{m}
+
\tilde{a}^{\frac{1}{2}} q^{|\tilde{s}|}
\frac{( 1 - \tilde{a}^{-1} q )}{( 1 - q )}
\sum_{m \geq 0}
\frac{( \tilde{a}^{-1} q^{2}, \tilde{a}^{-1} q^{3}; q^{2} )_{m}}{( q^{2}, q^{3}; q^{2} )_{m}} \left( \tilde{a} q^{2 |\tilde{s}|} \right)^{m} \notag \\ 
&=
\ {}_{2}\varphi_{1} ( \tilde{a}^{- 1} q^{2}, \tilde{a}^{- 1} q ; q ; q^{2}, \tilde{a} q^{2 |\tilde{s}|} )
+
\tilde{a}^{\frac{1}{2}} q^{|\tilde{s}|}
\frac{( \tilde{a}^{-1} q, q^{3}; q^{2} )_{\infty}}{( \tilde{a}^{-1} q^{3}, q; q^{2} )_{\infty}}
\ {}_{2}\varphi_{1} ( \tilde{a}^{- 1} q^{2}, \tilde{a}^{- 1} q^{3} ; q^{3} ; q^{2}, \tilde{a} q^{2 |\tilde{s}|} ), 
\label{b1qbinomial}
\end{align}
where we separate the summation over $j$ into the even-integer and the odd-integer part. Putting this back into the index, we have
\begin{align} 
\mathcal{I}_{{^{\text{X}}_{\text{Y}}}^{(+)}\text{Z}}
&=
q^{\frac{1}{8}} \tilde{a}^{- \frac{1}{4}}
\left( \tilde{a}^{- \frac{1}{2}} q^{\frac{1}{2}} \right)^{|\tilde{s}|}
\frac{( \tilde{a} ; q^{2} )_{\infty}}{( \tilde{a}^{- 1} q ; q^{2} )_{\infty}}
\left\{
{}_{2}\varphi_{1} ( \tilde{a}^{- 1} q^{2}, \tilde{a}^{- 1} q ; q ; q^{2}, \tilde{a} q^{2 |\tilde{s}|} )
\right. \notag \\ 
&\hspace{15em} +
\left.
\tilde{a}^{\frac{1}{2}} q^{|\tilde{s}|}
\frac{( \tilde{a}^{-1} q, q^{3}; q^{2} )_{\infty}}{( \tilde{a}^{-1} q^{3}, q; q^{2} )_{\infty}}
\ {}_{2}\varphi_{1} ( \tilde{a}^{- 1} q^{2}, \tilde{a}^{- 1} q^{3} ; q^{3} ; q^{2}, \tilde{a} q^{2 |\tilde{s}|} )
\right\} \notag \\ 
&=
q^{\frac{1}{8}}
\frac{( q^{2} ; q^{2} )_{\infty}}{( q ; q^{2} )_{\infty}}
\left\{
\tilde{a}^{- \frac{1}{4}}
\left( \tilde{a}^{- \frac{1}{2}} q^{\frac{1}{2}} \right)^{|\tilde{s}|}
\frac{( \tilde{a}, q ; q^{2} )_{\infty}}{( \tilde{a}^{- 1} q, q^{2} ; q^{2} )_{\infty}}
\ {}_{2}\varphi_{1} ( \tilde{a}^{- 1} q^{2}, \tilde{a}^{- 1} q ; q ; q^{2}, \tilde{a} q^{2 |\tilde{s}|} ) \right. \notag \\ 
&\hspace{7em} +
\left.
\tilde{a}^{\frac{1}{4}}
\left( \tilde{a}^{- \frac{1}{2}} q^{\frac{3}{2}} \right)^{|\tilde{s}|}
\frac{( \tilde{a}, q^{3}; q^{2} )_{\infty}}{( \tilde{a}^{-1} q^{3}, q^{2}; q^{2} )_{\infty}}
\ {}_{2}\varphi_{1} ( \tilde{a}^{- 1} q^{2}, \tilde{a}^{- 1} q^{3} ; q^{3} ; q^{2}, \tilde{a} q^{2 |\tilde{s}|} )
\right\}. 
\label{scib1xyz3}
\end{align}
Comparing \eqref{scib1xyz3} with \eqref{scib1sqed3}, we obtain the conclusion that $\mathcal{I}_{\text{SQED}^{( \mathcal{P} + )}} ( x, \alpha, B_{J} )$ \eqref{scib1sqed2} completely agrees with $\mathcal{I}_{{^{\text{X}}_{\text{Y}}}^{(+)}\text{Z}} ( x, \tilde{\alpha}, B_{V} )$ \eqref{scib1xyz2} under $\alpha = \tilde{\alpha}^{- 1}$ (equivalent to $a = \tilde{a}$) and $B_{J} = B_{V}$ (equivalent to $s = \tilde{s}$) as expected from the mirror map (Table \ref{mmap}). The crucial point is the division of the summation in the $q$-binomial theorem \eqref{b1qbinomial} which reproduces the even-holonomy and the odd-holonomy sector of the dynamical gauge field in \eqref{scib1sqed3}.

\subsection*{$\text{SQED}^{( \mathcal{P} - )}$ vs ${^{\text{X}}_{\text{Y}}}^{(-)}\text{Z}$}
We move to handling the example with the chosen background Wilson line phase $\theta_{J -}$ in the subsection \ref{Ptype}. The index of the SQED$^{( \mathcal{P} - )}$ is given by
\begin{align} 
&
\mathcal{I}_{\text{SQED}^{( \mathcal{P} - )}} ( x, \alpha, B_{J} ) \notag \\ 
&=
q^{\frac{1}{8}}
\frac{( q^{2} ; q^{2} )_{\infty}}{( q ; q^{2} )_{\infty}}
\oint_{C_0} \frac{d z}{2 \pi i z} z^{s}
\left\{
a^{- \frac{1}{4}}
\frac
{( z^{- 1} a^{\frac{1}{2}} q^{\frac{1}{2}}, z a^{\frac{1}{2}} q^{\frac{1}{2}}; q^{2} )_{\infty}}
{( z a^{- \frac{1}{2}} q^{\frac{1}{2}}, z^{- 1} a^{- \frac{1}{2}} q^{\frac{1}{2}} ; q^{2} )_{\infty}}
-
a^{\frac{1}{4}}
\frac
{( z^{- 1} a^{\frac{1}{2}} q^{\frac{3}{2}}, z a^{\frac{1}{2}} q^{\frac{3}{2}} ; q^{2} )_{\infty}}
{( z a^{- \frac{1}{2}} q^{\frac{3}{2}}, z^{- 1} a^{- \frac{1}{2}} q^{\frac{3}{2}} ; q^{2} )_{\infty}}
\right\}. 
\label{scib2sqed2}
\end{align}
This is the same index as the previous one except the weight in the summation over dynamical gauge field configurations. Namely, \eqref{scib2sqed2} can be evaluated on the residues of the poles \eqref{zpoles} as
\begin{align} 
&
\mathcal{I}_{\text{SQED}^{( \mathcal{P} - )}} \notag \\ 
&=
q^{\frac{1}{8}}
\frac{( q^{2} ; q^{2} )_{\infty}}{( q ; q^{2} )_{\infty}}
\left\{
a^{- \frac{1}{4}}
\left( a^{- \frac{1}{2}} q^{\frac{1}{2}} \right)^{| s |}
\frac{( a, q ; q^{2} )_{\infty}}{( a^{- 1} q, q^{2} ; q^{2} )_{\infty}}
\ {}_{2}\varphi_{1} ( a^{- 1} q^{2}, a^{- 1} q ; q ; q^{2}, a q^{2 |s|} )
\right. \notag \\ 
&\hspace{7em} -
\left.
a^{+ \frac{1}{4}} \left( a^{- \frac{1}{2}} q^{\frac{3}{2}} \right)^{| s |}
\frac{( a, q^{3} ; q^{2} )_{\infty}}{( a^{- 1} q^{3}, q^{2} ; q^{2} )_{\infty}}
\ {}_{2}\varphi_{1} ( a^{- 1} q^{2}, a^{- 1} q^{3} ; q^{3} ; q^{2}, a q^{2 |s|} )
\right\}. 
\label{scib2sqed3}
\end{align}

Now, we recall the index of ${^{\text{X}}_{\text{Y}}}^{(-)}\text{Z}$ shown in the subsection \ref{Ptype}
\begin{align} 
\mathcal{I}_{^{\text{X}}_{\text{Y}} \text{Z}^{(-)}} ( x, \tilde{\alpha}, B_{V} )
=
q^{\frac{1}{8}} \tilde{a}^{- \frac{1}{4}}
\left( \tilde{a}^{- \frac{1}{2}} q^{\frac{1}{2}} \right)^{|\tilde{s}|}
\frac{( - \tilde{a}^{- \frac{1}{2}} q^{1 + |\tilde{s}|} ; q )_{\infty}}{( - \tilde{a}^{\frac{1}{2}} q^{|\tilde{s}|} ; q )_{\infty}}
\frac{( \tilde{a} ; q^{2} )_{\infty}}{( \tilde{a}^{- 1} q ; q^{2} )_{\infty}}.
\label{scib2xyz2}
\end{align}
The way we follow to rewrite this index is nothing but the one that we done in the previous case, that is, we apply the $q$-binomial theorem \eqref{qbinomial} to the first fraction of \eqref{scib2xyz2},
\begin{align} 
&
\frac{( - \tilde{a}^{- \frac{1}{2}} q^{1 + |\tilde{s}|} ; q )_{\infty}}{( - \tilde{a}^{\frac{1}{2}} q^{|\tilde{s}|} ; q )_{\infty}} \notag \\ 
&=
{}_1 \varphi_0 ( \tilde{a}^{- 1} q ; - ; q, - \tilde{a}^{\frac{1}{2}} q^{|\tilde{s}|} ) \notag \\ 
&=
\sum_{n \geq 0} \frac{( \tilde{a}^{- 1} q; q )_{n}}{( q; q )_{n}} \left( - \tilde{a}^{\frac{1}{2}} q^{|\tilde{s}|} \right)^{n} \notag \\ 
&=
\sum_{m \geq 0}
\frac{( \tilde{a}^{- 1} q; q )_{2 m}}{( q; q )_{2 m}} \left( - \tilde{a}^{\frac{1}{2}} q^{|\tilde{s}|} \right)^{2 m}
+
\sum_{m \geq 0}
\frac{( \tilde{a}^{- 1} q; q )_{2 m + 1}}{( q; q )_{2 m + 1}} \left( - \tilde{a}^{\frac{1}{2}} q^{|\tilde{s}|} \right)^{2 m + 1} \notag \\ 
&=
\sum_{m \geq 0}
\frac{( \tilde{a}^{- 1} q, \tilde{a}^{- 1} q^{2}; q^{2} )_{m}}{( q, q^{2}; q^{2} )_{m}} \left( \tilde{a} q^{2 |\tilde{s}|} \right)^{m}
-
\tilde{a}^{\frac{1}{2}} q^{|\tilde{s}|}
\frac{( 1 - \tilde{a}^{- 1} q )}{( 1 - q )}
\sum_{m \geq 0}
\frac{( \tilde{a}^{- 1} q^{2}, \tilde{a}^{- 1} q^{3}; q^{2} )_{m}}{( q^{2}, q^{3}; q^{2} )_{m}} \left( \tilde{a} q^{2 |\tilde{s}|} \right)^{m} \notag \\ 
&=
{}_2 \varphi_1 ( \tilde{a}^{- 1} q, \tilde{a}^{- 1} q^{2}; q; q^{2}, \tilde{a} q^{2 |\tilde{s}|} )
-
\tilde{a}^{\frac{1}{2}} q^{|\tilde{s}|}
\frac{( \tilde{a}^{- 1} q, q^{3} ; q^{2})_{\infty}}{( q, \tilde{a}^{- 1} q^{3}; q^{2})_{\infty}}
{}_2 \varphi_1 ( \tilde{a}^{- 1} q^{2}, \tilde{a}^{- 1} q^{3}; q^{3}; q^{2}, \tilde{a} q^{2 |\tilde{s}|} ). 
\label{b2qbinomial}
\end{align}
Then, substituting it into \eqref{scib2xyz2} results in
\begin{align} 
\mathcal{I}_{^{\text{X}}_{\text{Y}} \text{Z}^{(-)}}
&=
q^{\frac{1}{8}} \tilde{a}^{- \frac{1}{4}}
\left( \tilde{a}^{- \frac{1}{2}} q^{\frac{1}{2}} \right)^{|\tilde{s}|}
\frac{( \tilde{a} ; q^{2} )_{\infty}}{( \tilde{a}^{- 1} q ; q^{2} )_{\infty}}
\left\{ {}_2 \varphi_1 ( \tilde{a}^{- 1} q, \tilde{a}^{- 1} q^{2}; q; q^{2}, \tilde{a} q^{2 |\tilde{s}|} ) \right. \notag \\ 
&\hspace{15em} -
\left.
\tilde{a}^{\frac{1}{2}} q^{|\tilde{s}|}
\frac{( \tilde{a}^{- 1} q, q^{3} ; q^{2})_{\infty}}{( q, \tilde{a}^{- 1} q^{3}; q^{2})_{\infty}}
{}_2 \varphi_1 ( \tilde{a}^{- 1} q^{2}, \tilde{a}^{- 1} q^{3}; q^{3}; q^{2}, \tilde{a} q^{2 |\tilde{s}|} ) \right\} \notag \\ 
&=
q^{\frac{1}{8}}
\frac{( q^{2}; q^{2} )_{\infty}}{( q; q^{2} )_{\infty}}
\left\{
\tilde{a}^{- \frac{1}{4}}
\left( \tilde{a}^{- \frac{1}{2}} q^{\frac{1}{2}} \right)^{|\tilde{s}|}
\frac{( \tilde{a}, q ; q^{2} )_{\infty}}{( \tilde{a}^{- 1} q, q^{2} ; q^{2} )_{\infty}}
{}_2 \varphi_1 ( \tilde{a}^{- 1} q, \tilde{a}^{- 1} q^{2}; q; q^{2}, \tilde{a} q^{2 |\tilde{s}|} ) \right. \notag \\ 
&\hspace{7em} -
\left.
\tilde{a}^{\frac{1}{4}}
\left( \tilde{a}^{- \frac{1}{2}} q^{\frac{3}{2}} \right)^{|\tilde{s}|}
\frac{( \tilde{a}, q^{3} ; q^{2})_{\infty}}{( \tilde{a}^{- 1} q^{3}, q^{2}; q^{2})_{\infty}}
{}_2 \varphi_1 ( \tilde{a}^{- 1} q^{2}, \tilde{a}^{- 1} q^{3}; q^{3}; q^{2}, \tilde{a} q^{2 |\tilde{s}|} ) \right\}. 
\label{scib2xyz3}
\end{align}
Comparing \eqref{scib2xyz3} with \eqref{scib2sqed3}, we reach equivalence between $\mathcal{I}_{\text{SQED}^{( \mathcal{P} - )}} ( x, \alpha, B_{J} )$ \eqref{scib2sqed2} and $\mathcal{I}_{{^{\text{X}}_{\text{Y}}}^{(-)}\text{Z}} ( x, \tilde{\alpha}, B_{V} )$ \eqref{scib2xyz2} under $\alpha = \tilde{\alpha}^{- 1}$ (equivalent to $a = \tilde{a}$) and $B_{J} = B_{V}$ (equivalent to $s = \tilde{s}$). As commented in $\mathcal{I}_{\text{SQED}^{( \mathcal{P} + )}} = \mathcal{I}_{{^{\text{X}}_{\text{Y}}}^{(+)}\text{Z}}$, dividing the summation in \eqref{b2qbinomial} correspondes to each holonomy sector of the dynamical gauge field with an appropriate phase.

\subsection*{$\text{SQED}^{( \mathcal{CP} + )}$ vs $\text{X}^{(+)}\text{Y}^{(+)}\text{Z}$}
We focus on mirror symmetry under new parity conditions explained in the subsection \ref{NewCP}. Here, we manifest the equality of the indices for mirror symmetry with the background gauge holonomy $B_{J}^{+}$ in the subsection \ref{CPtype}. The index of SQED$^{( \mathcal{CP} + )}$ is written by
\begin{align} 
&
\mathcal{I}_{\text{SQED}^{( \mathcal{CP} + )}} ( x, \alpha, w ) \notag \\ 
&=
\frac{1}{2}
q^{- \frac{1}{8}}
\frac{( q; q^{2} )_{\infty}}{( q^{2}; q^{2} )_{\infty}}
\sum_{m \in \mathbb{Z}}
\left( q^{\frac{1}{2}} a ^{- \frac{1}{2}} w \right)^{m}
\left[
\frac{( a^{- \frac{1}{2}} q^{m + 1}; q )_{\infty}}{( a^{\frac{1}{2}} q^{m}; q )_{\infty}}
+
\frac{( - a^{- \frac{1}{2}} q^{m + 1}; q )_{\infty}}{( - a^{\frac{1}{2}} q^{m}; q )_{\infty}}
\right]. 
\label{scib3sqed2}
\end{align}
Each term summed over monopole flux $m$ can be expressed by the general bilateral basic hypergeometric series \eqref{gbbhs} with $r = s = 1$ as
\begin{align} 
&
\mathcal{I}_{\text{SQED}^{( \mathcal{CP} + )}} \notag \\ 
&=
\frac{1}{2}
q^{- \frac{1}{8}}
\frac{( q; q^{2} )_{\infty}}{( q^{2}; q^{2} )_{\infty}}
\sum_{m \in \mathbb{Z}} \left( q^{\frac{1}{2}} a ^{- \frac{1}{2}} w \right)^{m}
\left[
\frac{( a^{- \frac{1}{2}} q; q )_{\infty}}{( a^{\frac{1}{2}}; q )_{\infty}}
\frac{( a^{\frac{1}{2}} ; q )_{m}}{( a^{- \frac{1}{2}} q ; q )_{m}}
+
\frac{( - a^{- \frac{1}{2}} q; q )_{\infty}}{( - a^{\frac{1}{2}}; q )_{\infty}}
\frac{( - a^{\frac{1}{2}} ; q )_{m}}{( - a^{- \frac{1}{2}} q ; q )_{m}} \right] \notag \\ 
&=
\frac{1}{2}
q^{- \frac{1}{8}}
\frac{( q; q^{2} )_{\infty}}{( q^{2}; q^{2} )_{\infty}}
\left[
\frac{( a^{- \frac{1}{2}} q; q )_{\infty}}{( a^{\frac{1}{2}}; q )_{\infty}}
{}_{1}\psi_{1} ( a^{\frac{1}{2}}; a^{- \frac{1}{2}} q; q, q^{\frac{1}{2}} a ^{- \frac{1}{2}} w )
+
\frac{( - a^{- \frac{1}{2}} q; q )_{\infty}}{( - a^{\frac{1}{2}}; q )_{\infty}}
{}_{1}\psi_{1} ( - a^{\frac{1}{2}}; - a^{- \frac{1}{2}} q; q, q^{\frac{1}{2}} a ^{- \frac{1}{2}} w ) \right]. 
\label{scib3sqed3}
\end{align}
Then, we translate $_{1}\psi_{1}$ into a certain combination of the $q$-shifted factorial by using the Ramanujan's sum \eqref{rama} as follows:
\begin{align} 
&
\mathcal{I}_{\text{SQED}^{( \mathcal{CP} + )}} \notag \\ 
&=
\frac{1}{2}
q^{- \frac{1}{8}}
\frac{( q; q^{2} )_{\infty}}{( q^{2}; q^{2} )_{\infty}}
\left[
\frac{( a^{- \frac{1}{2}} q; q )_{\infty}}{( a^{\frac{1}{2}}; q )_{\infty}}
\frac{( q, a^{- 1} q, w q^{\frac{1}{2}}, w^{- 1} q^{\frac{1}{2}}; q )_{\infty}}
{( a^{- \frac{1}{2}} q, a^{- \frac{1}{2}} q, a^{- \frac{1}{2}} w q^{\frac{1}{2}}, a^{- \frac{1}{2}} w^{- 1} q^{\frac{1}{2}}; q )_{\infty}} \right. \notag \\ 
&\hspace{12.25em} \left.
+
\frac{( - a^{- \frac{1}{2}} q; q )_{\infty}}{( - a^{\frac{1}{2}}; q )_{\infty}}
\frac{( q, a^{- 1} q, - w q^{\frac{1}{2}}, - w^{- 1} q^{\frac{1}{2}}; q )_{\infty}}
{( - a^{- \frac{1}{2}} q, - a^{- \frac{1}{2}} q, a^{- \frac{1}{2}} w q^{\frac{1}{2}}, a^{- \frac{1}{2}} w^{- 1} q^{\frac{1}{2}}; q )_{\infty}}
\right] \notag \\ 
&=
\frac{1}{2}
q^{- \frac{1}{8}}
\frac{( q; q^{2} )_{\infty}}{( q^{2}; q^{2} )_{\infty}}
\frac{( q, a^{- 1} q; q )_{\infty}}
{( a^{\frac{1}{2}}, a^{- \frac{1}{2}} q, - a^{\frac{1}{2}}, - a^{- \frac{1}{2}} q; q )_{\infty}}
\frac{1}
{( a^{- \frac{1}{2}} w q^{\frac{1}{2}}, a^{- \frac{1}{2}} w^{- 1} q^{\frac{1}{2}}; q )_{\infty}} \notag \\ 
&\hspace{7.5em} \times
\left[
( - a^{\frac{1}{2}}, - a^{- \frac{1}{2}} q, w q^{\frac{1}{2}}, w^{- 1} q^{\frac{1}{2}}; q )_{\infty}
+
( a^{\frac{1}{2}}, a^{- \frac{1}{2}} q, - w q^{\frac{1}{2}}, - w^{- 1} q^{\frac{1}{2}}; q )_{\infty}
\right] \notag \\ 
&=
\frac{1}{2}
q^{- \frac{1}{8}}
\frac{( q; q^{2} )_{\infty}}{( q^{2}; q^{2} )_{\infty}}
\frac{( q, q^{2}, a^{- 1} q, a^{- 1} q^{2}; q^{2} )_{\infty}}
{( a, a^{- 1} q^{2}; q^{2} )_{\infty}}
\frac{1}
{( a^{- \frac{1}{2}} w q^{\frac{1}{2}}, a^{- \frac{1}{2}} w^{- 1} q^{\frac{1}{2}}; q )_{\infty}} \notag \\ 
&\hspace{7.5em} \times
\left[
( - a^{\frac{1}{2}}, - a^{- \frac{1}{2}} q, w q^{\frac{1}{2}}, w^{- 1} q^{\frac{1}{2}}; q )_{\infty}
+
( a^{\frac{1}{2}}, a^{- \frac{1}{2}} q, - w q^{\frac{1}{2}}, - w^{- 1} q^{\frac{1}{2}}; q )_{\infty}
\right] \notag \\ 
&=
\frac{1}{2}
q^{- \frac{1}{8}}
\frac{( q, q, a^{- 1} q; q^{2} )_{\infty}}
{( a; q^{2} )_{\infty}}
\frac
{1}
{( a^{- \frac{1}{2}} w q^{\frac{1}{2}}, a^{- \frac{1}{2}} w q^{\frac{3}{2}}, a^{- \frac{1}{2}} w^{- 1} q^{\frac{1}{2}}, a^{- \frac{1}{2}} w^{- 1} q^{\frac{3}{2}}; q^{2} )_{\infty}}
\frac
{( a^{\frac{1}{2}} w^{- 1} q^{\frac{1}{2}}, a^{\frac{1}{2}} w q^{\frac{1}{2}}; q^{2} )_{\infty}}
{( a^{\frac{1}{2}} w^{- 1} q^{\frac{1}{2}}, a^{\frac{1}{2}} w q^{\frac{1}{2}}; q^{2} )_{\infty}}
\notag \\ 
&\hspace{7.5em} \times
\left[
( - a^{\frac{1}{2}}, - a^{- \frac{1}{2}} q, w q^{\frac{1}{2}}, w^{- 1} q^{\frac{1}{2}}; q )_{\infty}
+
( a^{\frac{1}{2}}, a^{- \frac{1}{2}} q, - w q^{\frac{1}{2}}, - w^{- 1} q^{\frac{1}{2}}; q )_{\infty}
\right] \notag \\ 
&=
\frac{1}{2}
q^{- \frac{1}{8}}
\frac{( a^{\frac{1}{2}} w^{- 1} q^{\frac{1}{2}}, a^{\frac{1}{2}} w q^{\frac{1}{2}}, a^{- 1} q; q^{2} )_{\infty}}
{( a^{- \frac{1}{2}} w q^{\frac{1}{2}}, a^{- \frac{1}{2}} w^{- 1} q^{\frac{1}{2}}, a; q^{2} )_{\infty}}
\frac{( q, q; q^{2} )_{\infty}}
{( a^{\frac{1}{2}} w^{- 1} q^{\frac{1}{2}}, a^{\frac{1}{2}} w q^{\frac{1}{2}}, a^{- \frac{1}{2}} w q^{\frac{3}{2}}, a^{- \frac{1}{2}} w^{- 1} q^{\frac{3}{2}}; q^{2} )_{\infty}} \notag \\ 
&\hspace{7.5em} \times
\left[
( - a^{\frac{1}{2}}, - a^{- \frac{1}{2}} q, w q^{\frac{1}{2}}, w^{- 1} q^{\frac{1}{2}}; q )_{\infty}
+
( a^{\frac{1}{2}}, a^{- \frac{1}{2}} q, - w q^{\frac{1}{2}}, - w^{- 1} q^{\frac{1}{2}}; q )_{\infty} \right], 
\label{scib3sqed4}
\end{align}
where we use the formulas \eqref{qdecomp} in the third equality. To make \eqref{scib3sqed4} easy to see, we define new variables as
\begin{align}
U = e^{2 \pi i z_{U}} := a^{\frac{1}{2}} q^{- \frac{1}{2}}, \hspace{2em}
W = e^{2 \pi i z_{W}} := w.
\label{newuw}
\end{align}
With respect to $U$ and $W$, the denominator of the second fraction in the first line of \eqref{scib3sqed4} becomes
\begin{align}
( a^{\frac{1}{2}} w^{- 1} q^{\frac{1}{2}}, a^{\frac{1}{2}} w q^{\frac{1}{2}}, a^{- \frac{1}{2}} w q^{\frac{3}{2}}, a^{- \frac{1}{2}} w^{- 1} q^{\frac{3}{2}}; q^{2} )_{\infty}
&=
( U W^{- 1} q, U W q, U^{- 1} W q, U^{- 1} W^{- 1} q; q^{2} )_{\infty} \notag \\ 
&=
( U W^{- 1} q, U^{- 1} W q; q^{2} )_{\infty}
( U W q, U^{- 1} W^{- 1} q; q^{2} )_{\infty} \notag \\ 
&=
\frac{\vartheta_{4} ( z_{U} - z_{W}, 2 \tau ) \vartheta_{4} ( z_{U} + z_{W}, 2 \tau )}
{( q^{2}, q^{2}; q^{2} )_{\infty}}, 
\label{theta4+-}
\end{align}
where we use the Jacobi's triple product identity for $\vartheta_{4}$ \eqref{triple4}. The Jacobi's triple product identities \eqref{triple4} and \eqref{triple3} can be also applied to the second line of \eqref{scib3sqed4} so that
\begin{align}
&
( - a^{\frac{1}{2}}, - a^{- \frac{1}{2}} q, w q^{\frac{1}{2}}, w^{- 1} q^{\frac{1}{2}}; q )_{\infty}
+
( a^{\frac{1}{2}}, a^{- \frac{1}{2}} q, - w q^{\frac{1}{2}}, - w^{- 1} q^{\frac{1}{2}}; q )_{\infty} \notag \\ 
&=
( - U q^{\frac{1}{2}}, - U^{- 1} q^{\frac{1}{2}}, W q^{\frac{1}{2}}, W^{- 1} q^{\frac{1}{2}}; q )_{\infty}
+
( U q^{\frac{1}{2}}, U^{- 1} q^{\frac{1}{2}}, - W q^{\frac{1}{2}}, - W^{- 1} q^{\frac{1}{2}}; q )_{\infty} \notag \\ 
&=
\frac{\vartheta_{3} ( z_{U}, \tau ) \vartheta_{4} ( z_{W}, \tau )}
{( q, q; q )_{\infty}}
+
\frac{\vartheta_{4} ( z_{U}, \tau ) \vartheta_{3} ( z_{W}, \tau )}
{( q, q; q )_{\infty}} \notag \\ 
&=
\frac
{1}
{( q, q; q )_{\infty}}
\left[
\vartheta_{3} ( z_{U}, \tau ) \vartheta_{4} ( z_{W}, \tau ) + \vartheta_{4} ( z_{U}, \tau ) \vartheta_{3} ( z_{W}, \tau )
\right]. 
\label{theta43+}
\end{align}
Substituting \eqref{theta4+-} and \eqref{theta43+} into the index \eqref{scib3sqed4}, we get
\begin{align} 
\mathcal{I}_{\text{SQED}^{( \mathcal{CP} + )}}
&=
\frac{1}{2}
q^{- \frac{1}{8}}
\frac
{( a^{\frac{1}{2}} w^{- 1} q^{\frac{1}{2}}, a^{\frac{1}{2}} w q^{\frac{1}{2}}, a^{- 1} q; q^{2} )_{\infty}}
{( a^{- \frac{1}{2}} w q^{\frac{1}{2}}, a^{- \frac{1}{2}} w^{- 1} q^{\frac{1}{2}}, a; q^{2} )_{\infty}}
\frac
{( q, q, q^{2}, q^{2}; q^{2} )_{\infty}}
{\vartheta_{4} ( z_{U} - z_{W}, 2 \tau ) \vartheta_{4} ( z_{U} + z_{W}, 2 \tau )} \notag \\ 
&\hspace{11.1em} \times
\frac
{1}
{( q, q; q )_{\infty}}
\left[
\vartheta_{3} ( z_{U}, \tau ) \vartheta_{4} ( z_{W}, \tau ) + \vartheta_{4} ( z_{U}, \tau ) \vartheta_{3} ( z_{W}, \tau )
\right] \notag \\ 
&=
\frac{1}{2}
q^{- \frac{1}{8}}
\frac
{( a^{\frac{1}{2}} w^{- 1} q^{\frac{1}{2}}, a^{\frac{1}{2}} w q^{\frac{1}{2}}, a^{- 1} q; q^{2} )_{\infty}}
{( a^{- \frac{1}{2}} w q^{\frac{1}{2}}, a^{- \frac{1}{2}} w^{- 1} q^{\frac{1}{2}}, a; q^{2} )_{\infty}}
\frac
{( q, q; q )_{\infty}}
{\vartheta_{4} ( z_{U} - z_{W}, 2 \tau ) \vartheta_{4} ( z_{U} + z_{W}, 2 \tau )} \notag \\ 
&\hspace{11.1em} \times
\frac
{1}
{( q, q; q )_{\infty}}
\left[
\vartheta_{3} ( z_{U}, \tau ) \vartheta_{4} ( z_{W}, \tau ) + \vartheta_{4} ( z_{U}, \tau ) \vartheta_{3} ( z_{W}, \tau )
\right] \notag \\ 
&=
\frac{1}{2}
q^{- \frac{1}{8}}
\frac
{( a^{\frac{1}{2}} w^{- 1} q^{\frac{1}{2}}, a^{\frac{1}{2}} w q^{\frac{1}{2}}, a^{- 1} q; q^{2} )_{\infty}}
{( a^{- \frac{1}{2}} w q^{\frac{1}{2}}, a^{- \frac{1}{2}} w^{- 1} q^{\frac{1}{2}}, a; q^{2} )_{\infty}}
\frac
{\vartheta_{3} ( z_{U}, \tau ) \vartheta_{4} ( z_{W}, \tau ) + \vartheta_{4} ( z_{U}, \tau ) \vartheta_{3} ( z_{W}, \tau )}
{\vartheta_{4} ( z_{U} - z_{W}, 2 \tau ) \vartheta_{4} ( z_{U} + z_{W}, 2 \tau )}. 
\label{scib3sqed5}
\end{align}
The product-to-sum identity \eqref{pts4443} tells us that the last fraction in \eqref{scib3sqed5} is just 2, that is,
\begin{align} 
\mathcal{I}_{\text{SQED}^{( \mathcal{CP} + )}} ( x, \alpha, w )
=
q^{- \frac{1}{8}}
\frac
{( a^{\frac{1}{2}} w^{- 1} q^{\frac{1}{2}}, a^{\frac{1}{2}} w q^{\frac{1}{2}}, a^{- 1} q; q^{2} )_{\infty}}
{( a^{- \frac{1}{2}} w q^{\frac{1}{2}}, a^{- \frac{1}{2}} w^{- 1} q^{\frac{1}{2}}, a; q^{2} )_{\infty}}.
\label{scib3sqed6}
\end{align}

Calling back the index of the X$^{( + )}$Y$^{( + )}$Z
\begin{align} 
\mathcal{I}_{\text{X}^{( + )}\text{Y}^{( + )}\text{Z}} ( x, \tilde{\alpha}, \tilde{w} )
&=
q^{- \frac{1}{8}}
\frac{( \tilde{a}^{\frac{1}{2}} \tilde{w}^{- 1} q^{\frac{1}{2}}, \tilde{a}^{\frac{1}{2}} \tilde{w} q^{\frac{1}{2}}, \tilde{a}^{- 1} q; q^{2} )_{\infty}}
{( \tilde{a}^{- \frac{1}{2}} \tilde{w} q^{\frac{1}{2}}, \tilde{a}^{- \frac{1}{2}} \tilde{w}^{- 1} q^{\frac{1}{2}}, \tilde{a}; q^{2} )_{\infty}},
\label{scib3xyz2}
\end{align}
we find that $\mathcal{I}_{\text{SQED}^{( \mathcal{CP} + )}} ( x, \alpha, w )$ \eqref{scib3sqed6} is perfectly compatible with $\mathcal{I}_{\text{X}^{( + )}\text{Y}^{( + )}\text{Z}} ( x, \tilde{\alpha}, \tilde{w} )$ \eqref{scib3xyz2} under identifications $\alpha = \tilde{\alpha}^{- 1}$ (equivalent to $a = \tilde{a}$) and $w = \tilde{w}$ (equivalent to $B_{J} = B_{V}$) which may be read off from the mirror map (Table \ref{mmap}).

\subsection*{$\text{SQED}^{( \mathcal{CP} - )}$ vs $\text{X}^{(-)}\text{Y}^{(-)}\text{Z}$}
Finally, we show mirror symmetry with the fixed background gauge holonomy $B_{J}^{-}$ in the subsection \ref{CPtype} where we provide the index of the SQED$^{( \mathcal{CP} - )}$ as
\begin{align} 
&
\mathcal{I}_{\text{SQED}^{( \mathcal{CP} - )}} ( x, \alpha, w ) \notag \\ 
&=
\frac{1}{2}
q^{- \frac{1}{8}}
\frac{( q; q^{2} )_{\infty}}{( q^{2}; q^{2} )_{\infty}}
\sum_{m \in \mathbb{Z}} \left( q^{\frac{1}{2}} a ^{- \frac{1}{2}} w \right)^{m}
\left[
\frac{( a^{- \frac{1}{2}} q^{m + 1}; q )_{\infty}}{( a^{\frac{1}{2}} q^{m}; q )_{\infty}}
-
\frac{( - a^{- \frac{1}{2}} q^{m + 1}; q )_{\infty}}{( - a^{\frac{1}{2}} q^{m}; q )_{\infty}}
\right]. 
\label{scib4sqed2}
\end{align}
As you can easily find, $\mathcal{I}_{\text{SQED}^{( \mathcal{CP} - )}}$ \eqref{scib4sqed2} differs from $\mathcal{I}_{\text{SQED}^{( \mathcal{CP} + )}}$ \eqref{scib3sqed2} by a relative sign originated from the BF term in summing up the dynamical gauge configuration. Thus, we rewrite \eqref{scib4sqed2} in the same manner that we done in the previous case: using the Ramanujan's sum \eqref{rama},
\begin{align}
&
\mathcal{I}_{\text{SQED}^{( \mathcal{CP} - )}} \notag \\ 
&=
\frac{1}{2}
q^{- \frac{1}{8}}
\frac
{( a^{\frac{1}{2}} w^{- 1} q^{\frac{3}{2}}, a^{\frac{1}{2}} w q^{\frac{3}{2}}, a^{- 1} q; q^{2} )_{\infty}}
{( a^{- \frac{1}{2}} w q^{\frac{3}{2}}, a^{- \frac{1}{2}} w^{- 1} q^{\frac{3}{2}}, a; q^{2} )_{\infty}}
\frac
{( q, q; q^{2} )_{\infty}}
{( a^{\frac{1}{2}} w^{- 1} q^{\frac{3}{2}}, a^{\frac{1}{2}} w q^{\frac{3}{2}}, a^{- \frac{1}{2}} w q^{\frac{1}{2}}, a^{- \frac{1}{2}} w^{- 1} q^{\frac{1}{2}}; q^{2} )_{\infty}} \notag \\ 
&\hspace{7.5em} \times
\left[
( - a^{\frac{1}{2}}, - a^{- \frac{1}{2}} q, w q^{\frac{1}{2}}, w^{- 1} q^{\frac{1}{2}}; q )_{\infty}
-
( a^{\frac{1}{2}}, a^{- \frac{1}{2}} q, - w q^{\frac{1}{2}}, - w^{- 1} q^{\frac{1}{2}}; q )_{\infty}
\right]. 
\label{scib4sqed3}
\end{align}
By means of the Jacobi's triple product identities \eqref{triple4} and \eqref{triple3}, the denominator of the second fraction in the first line of \eqref{scib4sqed3} is expressed in terms of $\vartheta_{4}$ as
\begin{align}
( a^{\frac{1}{2}} w^{- 1} q^{\frac{3}{2}}, a^{\frac{1}{2}} w q^{\frac{3}{2}}, a^{- \frac{1}{2}} w q^{\frac{1}{2}}, a^{- \frac{1}{2}} w^{- 1} q^{\frac{1}{2}}; q^{2} )_{\infty}
&=
( U W^{- 1} q^{2}, U W q^{2}, U^{- 1} W, U^{- 1} W^{- 1}; q^{2} )_{\infty} \notag \\ 
&=
( U W^{- 1} q^{2}, U^{- 1} W; q^{2} )_{\infty}
( U W q^{2}, U^{- 1} W^{- 1}; q^{2} )_{\infty} \notag \\ 
&=
\frac
{\vartheta_{4} ( z_{U} - z_{W} + \tau, 2 \tau ) \vartheta_{4} ( z_{U} + z_{W} + \tau, 2 \tau )}
{( q^{2}, q^{2}; q^{2} )_{\infty}}, 
\label{theta44+-tau}
\end{align}
and the second line of \eqref{scib4sqed3} turns to
\begin{align}
&
( - a^{\frac{1}{2}}, - a^{- \frac{1}{2}} q, w q^{\frac{1}{2}}, w^{- 1} q^{\frac{1}{2}}; q )_{\infty}
-
( a^{\frac{1}{2}}, a^{- \frac{1}{2}} q, - w q^{\frac{1}{2}}, - w^{- 1} q^{\frac{1}{2}}; q )_{\infty} \notag \\ 
&=
\frac
{1}
{( q, q; q )_{\infty}}
\left[
\vartheta_{3} ( z_{U}, \tau ) \vartheta_{4} ( z_{W}, \tau ) - \vartheta_{4} ( z_{U}, \tau ) \vartheta_{3} ( z_{W}, \tau )
\right], 
\label{theta43-}
\end{align}
where $( U, z_{U} )$ and $( W, z_{W} )$ are defined in \eqref{newuw}. Then, alternating \eqref{scib4sqed3} with \eqref{theta44+-tau} and \eqref{theta43-} leads to
\begin{align}
\mathcal{I}_{\text{SQED}^{( \mathcal{CP} - )}}
&=
\frac{1}{2}
q^{- \frac{1}{8}}
\frac
{( a^{\frac{1}{2}} w^{- 1} q^{\frac{3}{2}}, a^{\frac{1}{2}} w q^{\frac{3}{2}}, a^{- 1} q; q^{2} )_{\infty}}
{( a^{- \frac{1}{2}} w q^{\frac{3}{2}}, a^{- \frac{1}{2}} w^{- 1} q^{\frac{3}{2}}, a; q^{2} )_{\infty}}
\frac
{( q, q, q^{2}, q^{2}; q^{2} )_{\infty}}
{\vartheta_{4} ( z_{U} - z_{W} + \tau, 2 \tau ) \vartheta_{4} ( z_{U} + z_{W} + \tau, 2 \tau )} \notag \\ 
&\hspace{11.1em} \times
\frac
{1}
{( q, q; q )_{\infty}}
\left[
\vartheta_{3} ( z_{U}, \tau ) \vartheta_{4} ( z_{W}, \tau ) - \vartheta_{4} ( z_{U}, \tau ) \vartheta_{3} ( z_{W}, \tau )
\right] \notag \\ 
&=
\frac{1}{2}
q^{- \frac{1}{8}}
\frac
{( a^{\frac{1}{2}} w^{- 1} q^{\frac{3}{2}}, a^{\frac{1}{2}} w q^{\frac{3}{2}}, a^{- 1} q; q^{2} )_{\infty}}
{( a^{- \frac{1}{2}} w q^{\frac{3}{2}}, a^{- \frac{1}{2}} w^{- 1} q^{\frac{3}{2}}, a; q^{2} )_{\infty}}
\frac
{\vartheta_{3} ( z_{U}, \tau ) \vartheta_{4} ( z_{W}, \tau ) - \vartheta_{4} ( z_{U}, \tau ) \vartheta_{3} ( z_{W}, \tau )}
{\vartheta_{4} ( z_{U} - z_{W} + \tau, 2 \tau ) \vartheta_{4} ( z_{U} + z_{W} + \tau, 2 \tau )}. 
\label{scib4sqed4}
\end{align}
Furthermore, we apply the product-to-sum identity \eqref{pts4443} to the product of $\vartheta_{4}$'s in the denominator of \eqref{scib4sqed4} so that
\begin{align}
&
\vartheta_{4} ( z_{U} - z_{W} + \tau, 2 \tau ) \vartheta_{4} ( z_{U} + z_{W} + \tau, 2 \tau ) \notag \\ 
&=
\frac{1}{2}
\left\{
\vartheta_{4} ( z_{U} + \tau, \tau ) \vartheta_{3} ( z_{W}, \tau )
+
\vartheta_{3} ( z_{U} + \tau, \tau ) \vartheta_{4} ( z_{W}, \tau )
\right\} \notag \\ 
&=
\frac{1}{2}
\left\{
- q^{- \frac{1}{2}} U^{- 1}
\vartheta_{4} ( z_{U}, \tau ) \vartheta_{3} ( z_{W}, \tau )
+
q^{- \frac{1}{2}} U^{- 1}
\vartheta_{3} ( z_{U}, \tau ) \vartheta_{4} ( z_{W}, \tau )
\right\} \notag \\ 
&=
- \frac{1}{2} q^{- \frac{1}{2}} U^{- 1}
\left\{
\vartheta_{4} ( z_{U}, \tau ) \vartheta_{3} ( z_{W}, \tau )
-
\vartheta_{3} ( z_{U}, \tau ) \vartheta_{4} ( z_{W}, \tau )
\right\} \notag \\ 
&=
\frac{1}{2} a^{- \frac{1}{2}}
\left\{
\vartheta_{3} ( z_{U}, \tau ) \vartheta_{4} ( z_{W}, \tau )
-
\vartheta_{4} ( z_{U}, \tau ) \vartheta_{3} ( z_{W}, \tau )
\right\}, 
\end{align}
where we utilize the periodicity condition \eqref{period4} in the second equarity. As a result, we obtain
\begin{align}
\mathcal{I}_{\text{SQED}^{( \mathcal{CP} - )}} ( x, \alpha, w )
&=
q^{- \frac{1}{8}}
a^{\frac{1}{2}}
\frac
{( a^{\frac{1}{2}} w^{- 1} q^{\frac{3}{2}}, a^{\frac{1}{2}} w q^{\frac{3}{2}}, a^{- 1} q; q^{2} )_{\infty}}
{( a^{- \frac{1}{2}} w q^{\frac{3}{2}}, a^{- \frac{1}{2}} w^{- 1} q^{\frac{3}{2}}, a; q^{2} )_{\infty}}.
\label{scib4sqed5}
\end{align}

On the other hand, we retrieve the index of $\text{X}^{(-)}\text{Y}^{(-)}\text{Z}$
\begin{align} 
\mathcal{I}_{\text{X}^{(-)}\text{Y}^{(-)}\text{Z}} ( x, \tilde{\alpha}, \tilde{w} )
&=
q^{- \frac{1}{8}} \tilde{a}^{\frac{1}{2}}
\frac
{( \tilde{a}^{\frac{1}{2}} \tilde{w}^{- 1} q^{\frac{3}{2}}, \tilde{a}^{\frac{1}{2}} \tilde{w} q^{\frac{3}{2}}, \tilde{a}^{- 1} q; q^{2} )_{\infty}}
{( \tilde{a}^{- \frac{1}{2}} \tilde{w} q^{\frac{3}{2}}, \tilde{a}^{- \frac{1}{2}} \tilde{w}^{- 1} q^{\frac{3}{2}}, \tilde{a}; q^{2} )_{\infty}},
\label{scib4xyz2}
\end{align}
and conclude that mirror symmetry can be realized as $\mathcal{I}_{\text{SQED}^{( \mathcal{CP} - )}} ( x, \alpha, w ) = \mathcal{I}_{\text{X}^{(-)}\text{Y}^{(-)}\text{Z}} ( x, \tilde{\alpha}, \tilde{w} )$ with identifications $\alpha = \tilde{\alpha}^{- 1}$ (equivalent to $a = \tilde{a}$) and $w = \tilde{w}$ (equivalent to $B_{J} = B_{V}$).

\bibliographystyle{common_supp_files/sonota/utphys}
\bibliography{common_supp_files/Ref}

\bibliographystyle{sections/common_supp_files/sonota/utphys}
\bibliography{sections/common_supp_files/Ref}

\end{document}